\documentclass[hyperref,showpacs,showkeys,tightenlines,prd,12pt]{revtex4-2}
\usepackage{amsfonts}
\usepackage[debug,pdftex]{hyperref}
\usepackage{amsmath}
\usepackage{amssymb}
\usepackage{graphicx}
\usepackage{bibmods}
\usepackage{enumitem}
\usepackage[english]{babel}
\usepackage{blindtext}

\providecommand{\U}[1]{\protect\rule{.1in}{.1in}}

\begin{document}

\title{Strong First Order Phase Transition and $B$ Violation \\ in the Compact $341$ Model }
\author{\textbf{A. Boubakir}}
\email{alimaboubakir@yahoo.com}
\affiliation{Laboratoire de Physique Mathematique et Subatomique\\
Fr\`{e}res Mentouri Constantine1 University, Constantine, Algeria}
\author{\textbf{\ H. Aissaoui}}
\email{h.aissaoui@umc.edu.dz}
\affiliation{Laboratoire de Physique Mathematique et Subatomique\\
Fr\`{e}res Mentouri Constantine1 University, Constantine, Algeria}
\author{\textbf{\ N. Mebarki}}
\email{n.mebarki@umc.edu.dz}
\affiliation{Laboratoire de Physique Mathematique et Subatomique\\
Fr\`{e}res Mentouri Constantine1 University, Constantine, Algeria}

\begin{abstract}
Baryogenesis in the context of the compact $SU\left( 3\right) _{C}\otimes
SU\left( 4\right) _{L}\otimes U\left( 1\right) _{X}$ model is investigated.
Using the finite temperature effective potential approach together with
unitarity, stability and no ghost masses constraints, the existence of a
strong first order electroweak phase transition (EWPT) was shown and checked
numerically during all steps of the spontaneous breakdown of the gauge
symmetry of the model. Higgs masses regions fulfilling the EWPT criteria are
also discussed. Moreover, and as a byproduct of our study, the B-violation via sphaleron was also
emphasized.
\end{abstract}

\maketitle

\section{Introduction}


Electroweak baryogenesis (EWBG) remains a theoretically attractive and
experimentally testable scenario for explaining the cosmic baryon asymmetry. 
Particular attention is paid to Standard Model
extensions \cite{SM5}, \cite{baryo1}, \cite{baryo2}, \cite{baryo3} that may provide the necessary
ingredients for EWBG, and searches for the corresponding signatures at
high energy limits. Within the Standard Model (SM) \cite{SM}, 
\cite{St2}, \cite{St3}, \cite{St4}, \cite{SM1}, the EWBG cannot explain the
observed baryonic asymmetry of the universe. Indeed, the SM electroweak phase transition is of first
order (in order to have large deviations from thermal equilibrium) only if
the mass of the Higgs boson is less than $70$ GeV \cite{HB1}, \cite{HB2}. This
is in contradiction with the current experimental value which is around $125$
GeV \cite{HB4}. Moreover, the CP violation induced by the CKM phase does not
appear to be sufficient to generate the observed baryonic asymmetry \cite{CK1}, \cite%
{CK2}, \cite{CK3}. Thus, an extended SM theory is needed. One of such
possibilities is the widening of the gauge group symmetry leading to new
interactions and particle spectrum and which can be achieved at the TeV
scale, containing natural dark matter candidates \cite{DM1}, \cite{DM2} and
explaining the generation problem in the so-called $341$ model \cite{Mod341}, \cite{MMod341}, \cite{MMMod341}, \cite{MMMMod341}.

Electroweak phase transition (EWPT) is a type of symmetry breaking that plays
an important role at the early stage of the expanding universe where the scalar
potential is responsible for this. It is the transition between
symmetrical and asymmetrical phases, generating masses to elementary particles.


In order to describe the EWPT, it is better to use the technique of the
effective potential. It is a function containing the contributions coming from
fermions, bosons, and depends on temperature and vacuum expectation
values (VeVs) \cite{potef1}, \cite{potef2}, \cite{potef3}. It is worth
mentioning that the first order EWPT has  to be strong, that is,  the true vacuum expectation value (VeV)  $\upsilon _{c}$
has to be larger than the critical temperature, $\frac{\upsilon _{c}}{T_{c}}\geq 1$ (in the unit where Boltzmann's constant $k_B=1$) 
\cite{Strong cond}.

Among the extended models of our interest is the compact $341$ model, which
is based on the gauge symmetry group tensor product $SU\left( 3\right) _{C}\otimes
SU\left( 4\right) _{L}\otimes U\left( 1\right) _{X}$. In addition to the SM
particles spectra, the model contains twelve new gauge bosons, six exotic
quarks, two charged Higgses, a doubly charged Higgs and two neutral Higgses.
Some of the intriguing features of the $341$ model are the automatic existence of the
standard model Higgs, and the ability to contain a candidate
for cold dark matter \cite{DM1}, \cite{DM2}.

This paper is organized as follows, in Sec. \ref{sect2}, we briefly present the compact $
341$ model. In Sec. \ref{EPT}, we introduce the effective potential technique and show the structure of phase transition in 
the compact $341$ model. In Sec. \ref{sect4}, we present our numerical results taking into account the 
theoretical constraints imposed to the scalar potential.
In Sec. \ref{sect5}, we discuss the B-violation via the sphaleron approach. Finally, in Sec. \ref{sect6}, we draw our conclusions.

\section{The Compact $341$ Model}
\label{sect2}
\subsection{Particle content}


The compact $341$ model is described by the gauge group $SU\left( 3\right)
_{C}\otimes SU\left( 4\right) _{L}\otimes U\left( 1\right) _{X}$, and
contains all the particles of the SM with new gauge bosons, exotic quarks, and
three Higgs scalar quartets \cite{Mod341}, \cite{MMod341}. Like the SM, we
have three generations of fermions represented by the quartets:%
\begin{align}
\mathit{L}_{aL}& =\left( 
\begin{array}{c}
\upsilon _{a} \\ 
l_{a} \\ 
\upsilon _{a}^{c} \\ 
l_{a}^{c}%
\end{array}%
\right) _{L}\sim \left( 1,4,0\right),\text{ \ \ }a=e,\mu,\tau \\
Q_{1L}& =\left( 
\begin{array}{c}
u_{1} \\ 
d_{1} \\ 
U_{1} \\ 
J_{1}%
\end{array}%
\right) _{L}\sim \left( 3,4,2/3\right) \text{\ \, \ \ }\left\{ 
\begin{array}{c}
u_{1R}\sim \left( 3,3,1/3\right) \\ 
d_{1R}\sim \left( 3,1,-1/3\right) \\ 
U_{1R}\sim \left( 3,1,2/3\right) \\ 
J_{1R}\sim \left( 3,1,5/3\right)%
\end{array}%
\right.  \notag \\
Q_{iL}& =\left( 
\begin{array}{c}
d_{i} \\ 
u_{i} \\ 
D_{i} \\ 
J_{i}%
\end{array}%
\right) _{L}\sim \left( 3,4,-1/3\right),\left\{ 
\begin{array}{c}
u_{iR}\sim \left( 3,1,2/3\right) \\ 
d_{iR}\sim \left( 3,1,-1/3\right) \\ 
U_{iR}\sim \left( 3,1,-1/3\right) \\ 
J_{iR}\sim \left( 3,1,-4/3\right)%
\end{array}%
\right.,\text{ \ \ }i=2,3  \notag
\end{align}%
with $u_{1},$ $d_{1}$ are the up and down quarks, $U_{1},$ $J_{1},$ $J_{i},$ 
$D_{i}$ are the new exotic quarks with electric charges $2/3$, $5/3$, $-4/3,$
$-1/3$ respectively. We remind that the $U(1)$ charge $X$ is related to the
fermions electric charge by the relation:%
\begin{equation}
Q_{e}=(X\text{, }X-1\text{, }X\text{, }X+1)
\end{equation}%
The scalar sector contains three Higgs scalar quartets
\begin{align}
\eta & =\left( 
\begin{array}{c}
\eta _{1}^{0} \\ 
\eta _{1}^{-} \\ 
\eta _{2}^{0} \\ 
\eta _{2}^{+}
\end{array}
\right) =\left( 
\begin{array}{c}
\frac{1}{\sqrt{2}}(R_{\eta _{1}}+iI_{\eta _{1}}) \\ 
\eta _{1}^{-} \\ 
\frac{1}{\sqrt{2}}(\upsilon _{\eta }+R_{\eta _{2}}+iI_{\eta _{2}}) \\ 
\eta _{2}^{+}
\end{array}
\right) \sim \left( 1,4,0\right) \\
\rho & =\left( 
\begin{array}{c}
\rho _{1}^{+} \\ 
\rho ^{0} \\ 
\rho _{2}^{+} \\ 
\rho ^{++}
\end{array}
\right) =\left( 
\begin{array}{c}
\rho _{1}^{+} \\ 
\frac{1}{\sqrt{2}}(\upsilon _{\rho }+R_{\rho }+iI_{\rho }) \\ 
\rho _{2}^{+} \\ 
\rho ^{++}
\end{array}
\right) \sim \left( 1,4,1\right)  \notag \\
\chi & =\left( 
\begin{array}{c}
\chi _{1}^{-} \\ 
\chi ^{--} \\ 
\chi _{2}^{-} \\ 
\chi ^{0}%
\end{array}%
\right) =\left( 
\begin{array}{c}
\chi _{1}^{-} \\ 
\chi ^{--} \\ 
\chi _{2}^{-} \\ 
\frac{1}{\sqrt{2}}(\upsilon _{\chi }+R_{\chi }+iI_{\chi })%
\end{array}%
\right) \sim \left( 1,4,-1\right)  \notag
\end{align}%
the following neutral components develop three nontrivial vacuum expectation
values (VeVs)%
\begin{equation}
\eta =\frac{1}{\sqrt{2}}\left( 
\begin{array}{c}
0 \\ 
0 \\ 
\upsilon _{\eta } \\ 
0%
\end{array}%
\right),\text{ \ }\rho =\frac{1}{\sqrt{2}}\left( 
\begin{array}{c}
0 \\ 
\upsilon _{\rho } \\ 
0 \\ 
0%
\end{array}%
\right),\ \ \chi =\frac{1}{\sqrt{2}}\left( 
\begin{array}{c}
0 \\ 
0 \\ 
0 \\ 
\upsilon _{\chi }%
\end{array}%
\right)  \label{vevs}
\end{equation}%
with $R_{\eta _{1}},$ $R_{\eta _{2}},$ $R_{\rho },$ $R_{\chi }$ are the
CP-even scalar (real), $I_{\eta _{1}},$ $I_{\eta _{2}},$ $I_{\rho },$ $%
I_{\chi }$ are the CP-odd scalar (imaginary), the reason one chooses the eta
quadruplet to develop VeV only in the 3rd component is related to the fact
that we do not want to mix among ordinary and exotic quarks in the Yukawa
lagrangian, which guarantees the usual CKM mixing in the quark sector.
Equivalently, this is also possible if one also adds a new $Z_{3}$ discrete
symmetry to the model. The later will also allow for an appropriate
scenario for generating masses through effective operators. Spontaneous
symmetry breaking takes place in three different steps:

$\ast $ The first step :
\begin{equation}
SU\left( 4\right) _{L}\otimes U\left( 1\right) _{X}\text{\ }\underset{%
\upsilon _{\chi }}{\rightarrow }SU\left( 3\right) _{L}\otimes U\left(
1\right) _{X}
\end{equation}%

$\ast $ The second step :
\begin{equation}
SU\left( 3\right) _{L}\otimes U\left( 1\right) _{X}\text{\ }\underset{%
\upsilon _{\eta }}{\rightarrow }SU\left( 2\right) _{L}\otimes U\left(
1\right) _{Y}
\end{equation}%

$\ast $ The third step :
\begin{equation}
SU\left( 2\right) _{L}\otimes U\left( 1\right) _{Y}\underset{\upsilon _{\rho
}}{\rightarrow }U\left( 1\right) _{QED}
\end{equation}%
The VeVs $\upsilon _{\chi },$ $\upsilon _{\eta },$ $\upsilon _{\rho }$
satisfy the constraints:%
\begin{equation}
\upsilon _{\chi }>\upsilon _{\eta }>\upsilon _{\rho }
\end{equation}
So, in this model, there are two quite different scales of vacuum expectation
values: $\upsilon _{\eta }\sim O\left( \text{TeV}\right) $, $\upsilon _{\chi
}\sim O\left( \text{TeV}\right) $, and $\upsilon _{\rho }\approx 246$ GeV.
Following ref. \cite{MMod341}, the relationship between $SU(4)_L$ and $U(1)_X$ coupling constants $g_{L}$ and $g_{X }$ respectively is :
\begin{equation}
\frac{g_{X }^{2}}{g_{L}^{2}}=\frac{s_{w}^{2}}{1+4s_{w}^{2}}  \label{pp}
\end{equation}%
the relation (\ref{pp}) exhibits a landau pole when $s_{w}^{2}=\frac{1}{4}$
where $g_{X}\rightarrow \infty $ (comes infinite and $g_{L}$ finite) 
\cite{pol1}, \cite{pol2}, \cite{pol3}, The existence of a Landau pole for
the compact $341$ model at a scale of around $5$ TeV implies a natural cut-off for the model where 
one can circumvent the long standing hierarchy problem.

\subsection{The Higgs sector}


The scalar potential of the compact $341$ model \cite{MMod341} is given by%
\begin{align}
V\left( \eta,\rho,\chi \right) & =\mu _{\eta }^{2}\eta ^{+}\eta +\mu
_{\rho }^{2}\rho ^{+}\rho +\mu _{\chi }^{2}\chi ^{+}\chi  \label{POTEN} \\
& +\lambda _{1}\left( \eta ^{+}\eta \right) ^{2}+\lambda _{2}\left( \rho
^{+}\rho \right) ^{2}+\lambda _{3}\left( \chi ^{+}\chi \right) ^{2}  \notag
\\
& +\lambda _{4}\left( \eta ^{+}\eta \right) \left( \rho ^{+}\rho \right)
+\lambda _{5}\left( \eta ^{+}\eta \right) \left( \chi ^{+}\chi \right)
+\lambda _{6}\left( \rho ^{+}\rho \right) \left( \chi ^{+}\chi \right) 
\notag \\
& +\lambda _{7}\left( \rho ^{+}\eta \right) \left( \eta ^{+}\rho \right)
+\lambda _{8}\left( \chi ^{+}\eta \right) \left( \eta ^{+}\chi \right)
+\lambda _{9}\left( \rho ^{+}\chi \right) \left( \chi ^{+}\rho \right) +h.c 
\notag
\end{align}%
where $\lambda _{1},$ $\lambda _{2},$ $\lambda _{3},$ $\lambda _{4},$ $%
\lambda _{5},$ $\lambda _{6},$ $\lambda _{7},$ $\lambda _{8},$ $\lambda _{9}$
are dimensionless coupling constants, $\mu _{\eta,\rho,\chi }^{2}$ are the
mass dimension parameters satisfying the following relations when the
potential is minimized.These relations are given by the tadpole conditions :%
\begin{align}
\mu _{\eta }^{2}+\lambda _{1}v_{_{\eta }}^{2}+(\lambda _{4}v_{_{\rho
}}^{2}+\lambda _{5}v_{_{\chi }}^{2})/2& =0  \label{con} \\
\mu _{\rho }^{2}+\lambda _{2}v_{\rho }^{2}+(\lambda _{4}v_{_{\eta
}}^{2}+\lambda _{6}v_{_{\chi }}^{2})/2& =0  \notag \\
\mu _{\chi }^{2}+\lambda _{3}v_{_{\chi }}^{2}+(\lambda _{5}v_{_{\eta
}}^{2}+\lambda _{6}v_{_{\rho }}^{2})/2& =0  \notag
\end{align}%
The scalar potential depending on VeVs (\ref{POTEN}) can be written as
follows:%
\begin{align}
V\left( \eta,\rho,\chi \right) & =\mu _{\eta }^{2}\upsilon _{\eta
}^{2}/2+\mu _{\rho }^{2}\upsilon _{\rho }^{2}/2+\mu _{\chi }^{2}\upsilon
_{\chi }^{2}/2 \\
& +\lambda _{1}\upsilon _{\eta }^{2}/4+\lambda _{2}\upsilon _{\rho
}^{2}/4+\lambda _{3}\upsilon _{\chi }^{2}/4  \notag \\
& +\lambda _{4}\upsilon _{\eta }^{2}\upsilon _{\rho }^{2}/4+\lambda
_{5}\upsilon _{\eta }^{2}\upsilon _{\chi }^{2}/4+\lambda _{6}\upsilon _{\chi
}^{2}\upsilon _{\rho }^{2}/4  \notag
\end{align}%
Moreover, the CP-even elements of the mass matrix $\left( 4\times 4\right) $ in the
basis $(R_{\eta _{1}},$ $R_{\eta _{2}},$ $R_{\rho },$ $R_{\chi })$ are
written as
\begin{equation}
\left( 
\begin{array}{cccc}
0 & 0 & 0 & 0 \\ 
0 & 2\lambda _{1}v_{_{\eta }}^{2} & \lambda _{4}v_{_{\eta }}v_{_{\rho }} & 
\lambda _{5}v_{_{\eta }}v_{\chi } \\ 
0 & \lambda _{4}v_{_{\eta }}v_{_{\rho }} & 2\lambda _{2}v_{_{\rho }}^{2} & 
\lambda _{6}v_{_{\rho }}v_{\chi } \\ 
0 & \lambda _{5}v_{_{\eta }}v_{\chi } & \lambda _{6}v_{_{\rho }}v_{\chi } & 
2\lambda _{3}v_{_{\chi }}^{2}%
\end{array}%
\right)  \label{hh}
\end{equation}%
The eigenvalues of (\ref{hh}) are the masses of the neutral Higgses $H_{1}^{0}$%
, $H_{2}^{0}$, $H_{3}^{0}$ 
\begin{align}
M_{H_{1}^{0}}^{2}& =\left( \lambda _{2}+\frac{[\lambda _{3}\lambda
_{4}^{2}+\lambda _{6}(\lambda _{1}\lambda _{6}-\lambda _{4}\lambda _{5})]}{%
\lambda _{5}^{2}-4\lambda _{1}\lambda _{3}}\right) \upsilon _{_{\rho }}^{2}
\\
M_{H_{2}^{0}}^{2}& =\frac{1}{2}\left( \lambda _{1}+\lambda _{3}-\sqrt{%
(\lambda _{1}-\lambda _{3})^{2}+\lambda _{5}^{2}}\right) \upsilon _{\chi
}^{2}  \notag \\
M_{H_{3}^{0}}^{2}& =\frac{1}{2}\left( \lambda _{1}+\lambda _{3}+\sqrt{%
(\lambda _{1}-\lambda _{3})^{2}+\lambda _{5}^{2}}\right) \upsilon _{\chi
}^{2}  \notag \\
M_{G_{1}}^{2}& =0  \notag
\end{align}%
The CP-odd elements of the mass matrix $\left( 4\times 4\right) $ in the
basis $(I_{\eta _{1}},$ $I_{\eta _{2}},$ $I_{\rho },$ $I_{\chi })$ 
vanish and therefore, the neutral CP-odd Higgses are all massless%
\begin{equation}
M_{I_{\eta _{1}}}^{2}=M_{I_{\eta _{2}}}^{2}=0\text{, }M_{I_{\rho
}}^{2}=M_{I_{\chi }}^{2}=0
\end{equation}%
The mass matrices of the simply charged Higgs can be expressed according to
three basis:

1) In the basis $(\eta _{1}^{\pm },$ $\rho _{2}^{\pm })$ we have 
\begin{equation}
\frac{1}{2}\lambda _{7}\left( 
\begin{array}{cc}
\upsilon _{\eta }^{2} & \upsilon _{\rho }\upsilon _{\eta } \\ 
\upsilon _{\rho }\upsilon _{\eta } & \upsilon _{\rho }^{2}%
\end{array}%
\right)  \label{ch1}
\end{equation}%

with the eigenvalues 
\begin{equation}
M_{G_{1}^{\pm }}^{2}=0\text{, }M_{h_{1}^{\pm }}^{2}=\frac{1}{2}\lambda
_{7}(\upsilon _{\eta }^{2}+\upsilon _{\rho }^{2})
\end{equation}%

2) In the basis $(\eta _{2}^{\pm },$ $\chi _{2}^{\pm })$ we have 
\begin{equation}
\frac{1}{2}\lambda _{8}\left( 
\begin{array}{cc}
\upsilon _{\eta }^{2} & \upsilon _{\chi }\upsilon _{\eta } \\ 
\upsilon _{\chi }\upsilon _{\eta } & \upsilon _{\chi }^{2}%
\end{array}%
\right)  \label{ch 2}
\end{equation}%

and the eigenvalues 
\begin{equation}
M_{G_{2}^{\pm }}^{2}=0\text{, }M_{h_{2}^{\pm }}^{2}=\frac{1}{2}\lambda
_{8}(\upsilon _{\eta }^{2}+\upsilon _{\chi }^{2})
\end{equation}%

3) In the basis $(\rho _{1}^{\pm },$ $\chi _{1}^{\pm })$ we have%
\begin{equation}
M_{\rho _{1}^{\pm }}^{2}=0,\text{ }M_{\chi _{1}^{\pm }}^{2}=0
\end{equation}%
For the doubly charged Higgses, we have the mass matrix in the basis $%
(\chi ^{\pm \pm },$ $\rho ^{\pm \pm })$ 
\begin{equation}
\frac{1}{2}\lambda _{9}\left( 
\begin{array}{cc}
\upsilon _{\chi }^{2} & \upsilon _{\chi }\upsilon _{\rho } \\ 
\upsilon _{\chi }\upsilon _{\rho } & \upsilon _{\rho }^{2}%
\end{array}%
\right)  \label{ch D}
\end{equation}%
with the eigenvalues%
\begin{equation}
M_{G^{\pm \pm }}^{2}=0,\text{ }M_{h^{\pm \pm }}^{2}=\frac{1}{2}\lambda
_{9}(\upsilon _{\rho }^{2}+\upsilon _{\chi }^{2})
\end{equation}%
So, in this model, we have three neutral Higgses $(H_{1}^{0}$ the Higgs of the $%
SM)$ and three charged Higgses ($h_{1}^{\pm },$ $h_{2}^{\pm },$ $h^{\pm \pm }$%
), the Goldstone bosons $G^{\pm \pm }$ eaten by the doubly charged gauge
bosons $V^{\pm \pm },$ the Goldstone bosons $G_{1}^{\pm },$ $G_{2}^{\pm },$ $%
\rho _{1}^{\pm },$ $\chi _{1}^{\pm }$ eaten by the charged gauge bosons $%
W^{\pm },$ $K^{\pm },$ $Y^{\pm },$ $X^{\pm }$ and $I_{\eta _{1}},$ $I_{\eta
_{2}},$ $I_{\rho },$ $I_{\chi },$ $G_{1}$ eaten by the neutral gauge bosons $%
Z,$ $Z^{^{\prime }},$ $Z^{^{^{\prime \prime }}},$ $K,$ $K^{^{\prime }}$.

\subsection{The Fermions sector}

To obtain the fermion masses, we need the Yukawa interactions given by the
Lagrangian density \cite{Sphaleron331}: 
\begin{align}
\mathcal{L}_{Y}& =\lambda _{11}^{J}\overline{Q}_{1L}\chi J_{1R}+\lambda
_{ij}^{J}\overline{Q}_{iL}\chi ^{\ast }J_{jR}+\lambda _{11}^{U}\overline{Q}%
_{1L}\eta U_{1R}+\lambda _{ij}^{D}\overline{Q}_{iL}\eta ^{\ast }D_{jR}
\label{Yukw} \\
& +\lambda _{1a}^{d}\overline{Q}_{1L}\rho d_{aR}+\lambda _{ij}^{u}\overline{Q%
}_{iL}\rho ^{\ast }u_{aR}  \notag
\end{align}%
with $\lambda _{11}^{J},$ $\lambda _{ij}^{J},$ $\lambda _{11}^{U},$ $\lambda
_{ij}^{D}$ and $\lambda _{1a}^{d},$ $\lambda _{ij}^{u}$ are respectively the Yukawa
couplings of the exotic and the ordinary quarks. Like in
the $SM$, the masses of the usual quarks are proportional to the VeV $%
\upsilon _{\rho }$, they are involved in the third step of the spontanous
symmetry breaking (SSB) $SU\left( 2\right) _{L}\rightarrow U\left( 1\right)
_{Q\text{\ }}$. In what follows, we consider only the mass of the top quark:%
\begin{equation*}
m_{t}=\frac{\sqrt{2}}{2}\lambda _{33}^{u}\upsilon _{\rho }
\end{equation*}%
The masses of the exotic quarks $J_{1},$ $J_{2},$ $J_{3}$ are proportional
to the VeV $\upsilon _{\chi }$, and they are involved in the first step of
SSB $SU\left( 4\right) _{L}\rightarrow SU\left( 3\right) _{L}$, the mass of $%
J_{1}$ is:%
\begin{equation}
m_{J_{1}}=\frac{\sqrt{2}}{2}\lambda _{11}^{J}\upsilon _{\chi }
\end{equation}%
the mass matrix of $J_{2},$ $J_{3}$ in the basis $(J_{2},$ $J_{3})$ is
written as:%
\begin{equation}
\frac{\sqrt{2}}{2}\upsilon _{\chi }\left( 
\begin{array}{cc}
\lambda _{22}^{J} & \lambda _{23}^{J} \\ 
\lambda _{32}^{J} & \lambda _{33}^{J}%
\end{array}%
\right)
\end{equation}%
with the eigenvalues%
\begin{equation}
m_{J_{2}}=\frac{\sqrt{2}}{2}\lambda _{22}^{J}\upsilon _{\chi },\text{ }%
m_{J_{3}}=\frac{\sqrt{2}}{2}\lambda _{33}^{J}\upsilon _{\chi }
\end{equation}%
The masses of the exotic quarks $U_{1},$ $D_{2},$ $D_{3}$ are proportional
to the VeV $\upsilon _{\eta },$ and they are involved in the second step of
SSB $SU\left( 3\right) _{L}\rightarrow SU\left( 2\right) _{L}$, The mass of $%
U_{1}$ is:%
\begin{equation}
m_{U_{1}}=\frac{\sqrt{2}}{2}\lambda _{11}^{U}\upsilon _{\eta }
\end{equation}%
the mass matrix of $D_{2},$ $D_{3}$ in the basis $(D_{2},$ $D_{3})$ is
written as
\begin{equation}
\frac{\sqrt{2}}{2}\upsilon _{\eta }\left( 
\begin{array}{cc}
\lambda _{22}^{D} & \lambda _{23}^{D} \\ 
\lambda _{32}^{D} & \lambda _{33}^{D}%
\end{array}%
\right)
\end{equation}%
with the eigenvalues%
\begin{equation}
m_{D_{2}}=\frac{\sqrt{2}}{2}\lambda _{22}^{D}\upsilon _{\eta },\text{ }%
m_{D_{3}}=\frac{\sqrt{2}}{2}\lambda _{33}^{D}\upsilon _{\eta }\text{ }
\end{equation}%
A summary of the quarks masses formulation $m^{2}(\upsilon _{\eta },\upsilon
_{\rho },\upsilon _{\chi })=m^{2}(\upsilon _{\eta })+m^{2}(\upsilon _{\rho
})+m^{2}(\upsilon _{\chi })$ is shown in table 1:

\begin{center}
{\small Table 1: the quarks masses formulation}\\
$
\begin{tabular}{|l|l|l|l|l|}
\hline
$\text{quarks}$ & $m^{2}(\upsilon _{\eta },\upsilon _{\rho },\upsilon _{\chi
})$ & $m^{2}(\upsilon _{\eta })$ & $m^{2}(\upsilon _{\rho })$ & $%
m^{2}(\upsilon _{\chi })$ \\ \hline
$m_{D_{2}}^{2}$ & $\frac{1}{2}\lambda _{22}^{D^{2}}\upsilon _{\eta }^{2}$ & $%
\frac{1}{2}\lambda _{22}^{D^{2}}\upsilon _{\eta }^{2}$ & $0$ & $0$ \\ \hline
$m_{D_{3}}^{2}$ & $\frac{1}{2}\lambda _{33}^{D^{2}}\upsilon _{\eta }^{2}$ & $%
\frac{1}{2}\lambda _{33}^{D^{2}}\upsilon _{\eta }^{2}$ & $0$ & $0$ \\ \hline
$m_{U_{1}}^{2}$ & $\frac{1}{2}\lambda _{11}^{U^{2}}\upsilon _{\eta }^{2}$ & $%
\frac{1}{2}\lambda _{11}^{U^{2}}\upsilon _{\eta }^{2}$ & $0$ & $0$ \\ \hline
$m_{J_{2}}^{2}$ & $\frac{1}{2}\lambda _{22}^{J^{2}}\upsilon _{\chi }^{2}$ & $%
0$ & $0$ & $\frac{1}{2}\lambda _{22}^{J^{2}}\upsilon _{\chi }^{2}$ \\ \hline
$m_{J_{3}}^{2}$ & $\frac{1}{2}\lambda _{33}^{J^{2}}\upsilon _{\chi }^{2}$ & $%
0$ & $0$ & $\frac{1}{2}\lambda _{33}^{J^{2}}\upsilon _{\chi }^{2}$ \\ \hline
$m_{J_{1}}^{2}$ & $\frac{1}{2}\lambda _{11}^{J^{2}}\upsilon _{\chi }^{2}$ & $%
0$ & $0$ & $\frac{1}{2}\lambda _{11}^{J^{2}}\upsilon _{\chi }^{2}$ \\ \hline
$m_{t}^{2}$ & $\frac{1}{2}\lambda _{33}^{u^{2}}\upsilon _{\rho }^{2}$ & $0$
& $\frac{1}{2}\lambda _{33}^{u^{2}}\upsilon _{\rho }^{2}$ & $0$ \\ \hline
\end{tabular}
$
\end{center}

\subsection{The gauge Bosons}

Considering the Lagrangian density  $\mathcal{L}^{B}$ of the gauge bosons%
\begin{equation}
\mathcal{L}^{B}{\small =}\left( {\small D}_{\mu }\varkappa \right) ^{+}%
{\small (D}^{\mu }{\small \varkappa )+}\left( {\small D}_{\mu }{\small \eta }%
\right) ^{{\small +}}{\small (D}^{\mu }{\small \eta )+}\left( {\small D}%
_{\mu }{\small \rho }\right) ^{{\small +}}{\small (D}^{\mu }{\small \rho )}
\end{equation}%
where $D_{\mu }$ is the covariant derivative given by%
\begin{equation}
{\small D}^{\mu }{\small =\partial }^{\mu }{\small -}\frac{{\small ig}_{_{L}}%
}{{\small 2}}{\small \lambda }_{\alpha }{\small A}_{\alpha }^{\mu }{\small %
-ig}_{_{x}}{\small XB}^{\mu }{\small =\partial }^{\mu }{\small -iP}^{\mu }
\end{equation}%
and
\begin{equation}
{\small P}^{\mu }{\small =}\frac{g_{_{L}}}{2}\left( 
\begin{array}{cccc}
\begin{array}{c}
{\small W}_{3}^{\mu }{\small +W}_{8}^{\mu }{\small /}\sqrt{3} \\ 
{\small +W}_{15}^{\mu }{\small /}\sqrt{6} \\ 
{\small +2}\frac{g_{_{x}}}{g_{_{L}}}{\small XB}^{\mu }%
\end{array}
& {\small (W}_{1}^{^{\mu }}{\small +iW}_{2}^{^{\mu }}{\small )} & {\small (W}%
_{6}^{^{\mu }}{\small +iW}_{7}^{^{\mu }}{\small )} & {\small (W}_{13}^{^{\mu
}}{\small +iW}_{14}^{^{\mu }}{\small )} \\ 
{\small (W}_{1}^{^{\mu }}{\small -iW}_{2}^{^{\mu }}{\small )} & 
\begin{array}{c}
{\small -W}_{3}^{\mu }{\small +W}_{8}^{\mu }{\small /}\sqrt{3} \\ 
{\small +W}_{15}^{\mu }{\small /}\sqrt{6} \\ 
{\small +2}\frac{g_{_{x}}}{g_{_{L}}}{\small XB}^{\mu }%
\end{array}
& {\small (W}_{4}^{^{\mu }}{\small +iW}_{5}^{^{\mu }}{\small )} & {\small (W}%
_{9}^{^{\mu }}{\small +iW}_{10}^{^{\mu }}{\small )} \\ 
{\small (W}_{1}^{^{\mu }}{\small -iW}_{2}^{^{\mu }}{\small )} & {\small (W}%
_{4}^{^{\mu }}{\small -iW}_{5}^{^{\mu }}{\small )} & 
\begin{array}{c}
{\small -2W}_{8}^{\mu }{\small /}\sqrt{3} \\ 
{\small +W}_{15}^{\mu }{\small /}\sqrt{6} \\ 
{\small +2}\frac{g_{_{x}}}{g_{_{L}}}{\small XB}^{\mu }%
\end{array}
& {\small (W}_{11}^{^{\mu }}{\small +iW}_{12}^{^{\mu }}{\small )} \\ 
{\small (W}_{13}^{^{\mu }}{\small -iW}_{14}^{^{\mu }}{\small )} & {\small (W}%
_{9}^{^{\mu }}{\small -iW}_{10}^{^{\mu }}{\small )} & {\small (W}%
_{11}^{^{\mu }}{\small -iW}_{12}^{^{\mu }}{\small )} & 
\begin{array}{c}
{\small -3W}_{15}^{\mu }{\small /}\sqrt{6} \\ 
{\small +2}\frac{g_{_{x}}}{g_{_{L}}}{\small XB}^{\mu }%
\end{array}%
\end{array}%
\right)
\end{equation}
$W_{i}^{\mu }(i=1....15)$ and $B_{\mu }$ are the gauge fields.

For the charged gauge bosons, in the basis $(W_{1}^{^{\mu }},$ $W_{2}^{^{\mu
}}),(W_{4}^{^{\mu }},$ $W_{5}^{^{\mu }}),$ $(W_{6}^{^{\mu }},$ $W_{7}^{^{\mu
}}),$ $(W_{9}^{^{\mu }},$ $W_{10}^{^{\mu }}),$ $(W_{11}^{^{\mu }},$ $%
W_{12}^{^{\mu }})$,$(W_{13}^{^{\mu }},$ $W_{14}^{^{\mu }})$ one has the
following mass matrices:%
\begin{eqnarray}
&&\frac{{\small g}_{_{L}}^{2}}{{\small 4}}\left( 
\begin{array}{cc}
{\small \upsilon }_{\rho }^{2} & {\small 0} \\ 
{\small 0} & {\small \upsilon }_{\rho }^{2}%
\end{array}%
\right) {\small,}\text{ }\frac{{\small g}_{_{L}}^{2}}{{\small 4}}\left( 
\begin{array}{cc}
{\small (\upsilon }_{\rho }^{2}{\small +\upsilon }_{\eta }^{2}{\small )} & 
{\small 0} \\ 
{\small 0} & {\small (\upsilon }_{\rho }^{2}{\small +\upsilon }_{\eta }^{2}%
{\small )}%
\end{array}%
\right) \\
&&\frac{{\small g}_{_{L}}^{2}}{{\small 4}}\left( 
\begin{array}{cc}
{\small \upsilon }_{\eta }^{2} & {\small 0} \\ 
{\small 0} & {\small \upsilon }_{\eta }^{2}%
\end{array}%
\right),\text{ }\frac{{\small g}_{_{L}}^{2}}{{\small 4}}\left( 
\begin{array}{cc}
{\small \upsilon }_{\rho }^{2}{\small +\upsilon }_{\chi }^{2} & {\small 0}
\\ 
{\small 0} & {\small \upsilon }_{\rho }^{2}{\small +\upsilon }_{\chi }^{2}%
\end{array}%
\right) \\
&&\frac{{\small g}_{_{L}}^{2}}{{\small 4}}\left( 
\begin{array}{cc}
{\small \upsilon }_{\eta }^{2}{\small +\upsilon }_{\chi }^{2} & {\small 0}
\\ 
{\small 0} & {\small \upsilon }_{\eta }^{2}{\small +\upsilon }_{\chi }^{2}%
\end{array}%
\right),\text{ }\frac{{\small g}_{_{L}}^{2}}{{\small 4}}\left( 
\begin{array}{cc}
{\small \upsilon }_{\chi }^{2} & {\small 0} \\ 
{\small 0} & {\small \upsilon }_{\chi }^{2}%
\end{array}%
\right)  \notag
\end{eqnarray}%
To obtain the masses:%
\begin{eqnarray}
{\small M}_{W^{\pm }}^{2} &{\small =}&{\small g}_{_{L}}^{2}{\small \upsilon }%
_{\rho }^{2}{\small /4,}\text{ }{\small M}_{^{K_{1}^{\pm }}}^{2}{\small =g}%
_{_{L}}^{2}{\small (\upsilon }_{\rho }^{2}{\small +\upsilon }_{\eta }^{2}%
{\small )/4} \\
{\small M}_{K,\text{ }K^{^{\prime }}}^{2} &{\small =}&\frac{{\small g}%
_{_{L}}^{2}}{{\small 4}}{\small \upsilon }_{\eta }^{2}{\small,}\text{ }%
{\small M}_{Y^{\pm }}^{2}{\small =}\frac{{\small g}_{_{L}}^{2}{\small %
(\upsilon }_{\rho }^{2}{\small +\upsilon }_{\chi }^{2}{\small )}}{{\small 4}}
\notag \\
{\small M}_{V^{\pm \pm }}^{2} &{\small =}&\frac{{\small g}_{_{L}}^{2}{\small %
(\upsilon }_{\eta }^{2}{\small +\upsilon }_{\chi }^{2}{\small )}}{{\small 4}}%
{\small,}\text{ }{\small M}_{X^{\pm }}^{2}{\small =}\frac{{\small g}%
_{_{L}}^{2}{\small (\upsilon }_{\chi }^{2}{\small )}}{{\small 4}}  \notag
\end{eqnarray}%
respectively. For the neutral bosons in the basis $(W_{3}^{^{\mu }},$ $
W_{8}^{^{\mu }},$ $W_{15}^{^{\mu }},$ $B^{\mu })$ we have the following mass
matrix:%
\begin{equation}
\frac{g_{_{L}}^{2}}{4}\left( 
\begin{array}{cccc}
{\small \upsilon }_{\rho }^{2} & \frac{{\small -\upsilon }_{\rho }^{2}}{%
\sqrt{3}} & {\small -}\frac{{\small \upsilon }_{\rho }^{2}}{\sqrt{6}} & 
{\small -2X}\frac{g_{_{x}}}{g_{_{L}}}{\small \upsilon }_{\rho }^{2} \\ 
{\small -}\frac{{\small \upsilon }_{\rho }^{2}}{\sqrt{3}} & \frac{{\small %
(4\upsilon }_{\eta }^{2}{\small +\upsilon }_{\rho }^{2}{\small )}}{3} & 
\frac{{\small (\upsilon }_{\rho }^{2}{\small -2\upsilon }_{\eta }^{2}{\small %
)}}{3\sqrt{2}} & {\small 2X}\frac{g_{_{x}}}{g_{_{L}}}\frac{{\small (\upsilon 
}_{\rho }^{2}{\small -2\upsilon }_{\eta }^{2}{\small )}}{\sqrt{3}} \\ 
{\small -}\frac{{\small \upsilon }_{\rho }^{2}}{\sqrt{6}} & \frac{{\small %
(\upsilon }_{\rho }^{2}{\small -2\upsilon }_{\eta }^{2}{\small )}}{3\sqrt{2}}
& \frac{{\small (\upsilon }_{\eta }^{2}{\small +\upsilon }_{\rho }^{2}%
{\small +9\upsilon }_{\chi }^{2}{\small )}}{6} & {\small 2X}\frac{g_{_{x}}}{%
g_{_{L}}}\frac{{\small (\upsilon }_{\eta }^{2}{\small +\upsilon }_{\rho }^{2}%
{\small -3\upsilon }_{\chi }^{2}{\small )}}{\sqrt{6}} \\ 
{\small -2X}\frac{g_{_{x}}}{g_{_{L}}}{\small \upsilon }_{\rho }^{2} & 
{\small 2X}\frac{g_{_{x}}}{g_{_{L}}}{\small (\upsilon }_{\rho }^{2}{\small %
-2\upsilon }_{\eta }^{2}{\small )/}\sqrt{3} & {\small 2X}\frac{g_{_{x}}}{%
g_{_{L}}}\frac{{\small (\upsilon }_{\eta }^{2}{\small +\upsilon }_{\rho }^{2}%
{\small -3\upsilon }_{\chi }^{2}{\small )}}{\sqrt{6}} & {\small 4X}^{2}\frac{%
g_{_{x}}^{2}}{g_{_{L}}^{2}}{\small (\upsilon }_{\eta }^{2}{\small +\upsilon }%
_{\rho }^{2}{\small +3\upsilon }_{\chi }^{2}{\small )}%
\end{array}%
\right)
\end{equation}%
and therefore, the masses of $\gamma,$ $Z,$ $Z^{^{\prime }},$ $%
Z^{^{^{\prime \prime }}}$ are:%
\begin{align}
{\small M}_{\gamma }^{2}& {\small =0,}\text{ \ }{\small M}_{Z}^{2}{\small =g}%
_{_{L}}^{2}{\small \upsilon }_{\rho }^{2}{\small /4}\cos ^{2}{\small \theta }%
_{W}{\small,}\text{ \ }{\small M}_{Z^{^{\prime }}}^{2}{\small =g}_{_{L}}^{2}%
{\small (}\cos ^{2}{\small \theta }_{W}{\small \upsilon }_{\eta }^{2}{\small %
)/(3-4}\sin ^{2}{\small \theta }_{W}{\small )} \\
{\small M}_{Z^{^{^{\prime \prime }}}}^{2}& {\small =g}_{_{L}}^{2}{\small %
\upsilon }_{\eta }^{2}{\small [(1-4}\sin ^{2}{\small \theta }_{W}{\small %
)+(3-4}\sin ^{2}{\small \theta }_{W}{\small )]/8(3-4}\sin ^{2}{\small \theta 
}_{W}{\small )(1-4}\sin ^{2}{\small \theta }_{W}{\small )}  \notag
\end{align}%
Here $\theta _{W}$ is the Weinberg angle. Table 2 summarizes the gauge
bosons and Higgs masses formulation $m^{2}(\upsilon _{\eta },\upsilon _{\rho
},\upsilon _{\chi })=m^{2}(\upsilon _{\eta })+m^{2}(\upsilon _{\rho
})+m^{2}(\upsilon _{\chi })$:

\begin{center}
\small{Table 2 : The gauge bosons and Higgs masses formulation}
$
\begin{tabular}{|c|c|c|c|c|}
\hline
$\small\text{Bosons}$ & ${\small m}^{2}{\small (\upsilon }_{\eta }{\small %
,\upsilon }_{\rho }{\small,\upsilon }_{\chi }{\small )}$ & ${\small m}^{2}%
{\small (\upsilon }_{\eta }{\small )}$ & ${\small m}^{2}{\small (\upsilon }%
_{\rho }{\small )}$ & ${\small m}^{2}{\small (\upsilon }_{\chi }{\small )}$
\\ \hline
${\small m}_{W^{\pm }}^{2}$ & $\frac{g_{L}^{2}\upsilon _{\rho }^{2}}{4}$ & $%
{\small 0}$ & $\small{(80.40 \ GeV)^{2}}$ & ${\small 0}$ \\ \hline
${\small m}_{K_{1}^{\pm }}^{2}$ & $\frac{g_{L}^{2}(\upsilon _{\rho
}^{2}+\upsilon _{\eta }^{2})}{4}$ & $\frac{g_{L}^{2}\upsilon _{\eta }^{2}}{4}
$ & $\frac{g_{L}^{2}\upsilon _{\rho }^{2}}{4}$ & ${\small 0}$ \\ \hline
${\small m}_{K}^{2}$ & $\frac{g_{L}^{2}\upsilon _{\eta }^{2}}{4}$ & $\frac{%
g_{L}^{2}\upsilon _{\eta }^{2}}{4}$ & ${\small 0}$ & ${\small 0}$ \\ \hline
${\small m}_{V^{\pm \pm }}^{2}$ & $\frac{g_{L}^{2}(\upsilon _{\rho
}^{2}+\upsilon _{\chi }^{2})}{4}$ & ${\small 0}$ & $\frac{g_{L}^{2}\upsilon
_{\rho }^{2}}{4}$ & $\frac{g_{L}^{2}\upsilon _{\chi }^{2}}{4}$ \\ \hline
${\small m}_{Y^{\pm }}^{2}$ & $\frac{g_{L}^{2}(\upsilon _{\eta
}^{2}+\upsilon _{\chi }^{2})}{4}$ & $\frac{g_{L}^{2}\upsilon _{\eta }^{2}}{4}
$ & ${\small 0}$ & $\frac{g_{L}^{2}\upsilon _{\chi }^{2}}{4}$ \\ \hline
${\small m}_{X^{\pm }}^{2}$ & $\frac{g_{L}^{2}\upsilon _{\chi }^{2}}{4}$ & $%
{\small 0}$ & ${\small 0}$ & $\frac{g_{L}^{2}\upsilon _{\chi }^{2}}{4}$ \\ 
\hline
${\small m}_{Z}^{2}$ & $\frac{g_{_{L}}^{2}\upsilon _{\rho }^{2}}{4\cos
^{2}\theta _{W}\text{ }}$ & ${\small 0}$ & $(91.68${\small \ GeV}$)^{2}$ & $%
{\small 0}$ \\ \hline
${\small m}_{Z^{^{\prime }}}^{2}$ & $\frac{g_{_{L}}^{2}\text{ }\cos
^{2}\theta _{W}\text{ }\upsilon _{\eta }^{2}}{(3-4\sin ^{2}\theta _{W})}$ & $%
\frac{g_{_{L}}^{2}\text{ }\cos ^{2}\theta _{W}\text{ }\upsilon _{\eta }^{2}}{%
(3-4\sin ^{2}\theta _{W})}$ & ${\small 0}$ & ${\small 0}$ \\ \hline
${\small m}_{Z^{^{\prime \prime }}}^{2}$ & $\frac{g_{_{L}}^{2}\text{ }%
\upsilon _{\eta }^{2}[(1-4\sin ^{2}\theta _{W}\text{ })+(3-4\sin ^{2}\theta
_{W})]}{8(3-4\sin ^{2}\theta _{W})(1-4\sin ^{2}\theta _{W}\text{ })}$ & $%
\frac{g_{_{L}}^{2}\text{ }\upsilon _{\eta }^{2}[(1-4\sin ^{2}\theta _{W}%
\text{ })+(3-4\sin ^{2}\theta _{W})]}{8(3-4\sin ^{2}\theta _{W})(1-4\sin
^{2}\theta _{W}\text{ })}$ & ${\small 0}$ & ${\small 0}$ \\ \hline
${\small m}_{h_{1}^{\pm }}^{2}$ & $\frac{\lambda _{7}(\upsilon _{\eta
}^{2}+\upsilon _{\rho }^{2})}{2}$ & $\frac{\lambda _{7}\upsilon _{\eta }^{2}%
}{2}$ & $\frac{\lambda _{7}\upsilon _{\rho }^{2}}{2}$ & ${\small 0}$ \\ 
\hline
${\small m}_{h_{2}^{\pm }}^{2}$ & $\frac{\lambda _{8}(\upsilon _{\eta
}^{2}+\upsilon _{\chi }^{2})}{2}$ & $\frac{\lambda _{8}\upsilon _{\eta }^{2}%
}{2}$ & ${\small 0}$ & $\frac{\lambda _{8}\upsilon _{\chi }^{2}}{2}$ \\ 
\hline
${\small m}_{h^{\pm \pm }}^{2}$ & $\frac{\lambda _{9}(\upsilon _{\rho
}^{2}+\upsilon _{\chi }^{2})}{2}$ & ${\small 0}$ & $\frac{\lambda
_{9}\upsilon _{\rho }^{2}}{2}$ & $\frac{\lambda _{9}\upsilon _{\chi }^{2}}{2}
$ \\ \hline
${\small m}_{H_{1}^{0}}^{2}$ & ${\small (\lambda }_{2}{\small +}\frac{%
[\lambda _{3}\lambda _{4}^{2}+\lambda _{6}(\lambda _{1}\lambda _{6}-\lambda
_{4}\lambda _{5})]}{\lambda _{5}^{2}-4\lambda _{1}\lambda _{3}}{\small %
)\upsilon }_{_{\rho }}^{2}$ & ${\small 0}$ & $(125${\small GeV}$)^{2}$ & $%
{\small 0}$ \\ \hline
${\small m}_{H_{2}^{0}}^{2}$ & $\frac{\left( \lambda _{1}+\lambda _{3}-\sqrt{%
(\lambda _{1}-\lambda _{3})^{2}+\lambda _{5}^{2}}\right) \upsilon _{\chi
}^{2}}{2}$ & ${\small 0}$ & ${\small 0}$ & $(750${\small \ GeV}$)^{2}$ \\ 
\hline
${\small m}_{H_{3}^{0}}^{2}$ & $\frac{\left( \lambda _{1}+\lambda _{3}+\sqrt{%
(\lambda _{1}-\lambda _{3})^{2}+\lambda _{5}^{2}}\right) {\small %
\upsilon }_{\chi }^{2}}{2}$ & ${\small 0}$ & ${\small 0}$ & $\frac{\left(
\lambda _{1}+\lambda _{3}+\sqrt{(\lambda _{1}-\lambda _{3})^{2}+\lambda
_{5}^{2}}\right){\small \upsilon }_{\chi }^{2} }{2}$ \\ \hline
\end{tabular}
$ 
\end{center}

\section{Electroweak Phase Transition}
\label{EPT}
\subsection{Effective Potential}

In a perturbative analysis of the electroweak phase transition, the important
tool is the effective potentiel at finite temperature \cite{28arti3}, \cite{effec}, 
\cite{potef4}, \cite{potef5}. It is the contribution coming from fermions,
bosons and Higgses. This function also depends on the VeV's and temperature and is 
given at one loop order by \cite{8arti3}
\begin{equation}
V_{eff}\left( \phi,T\right) =V_{0}\left( \phi \right) +V_{1}\left( \phi
\right) +\Delta V_{1}^{(T)}\left( \phi,T\right)  \label{poteff}
\end{equation}%
where $V_{0}$ and $V_{1}$ are respectively the tree-level and the one-loop effective
potential at $T=0$. The third term $\Delta V_{1}^{(T)}\left(
\phi,T\right) $ contains the one loop thermal corrections. The expression
of $V_{1}\left( \phi \right) $ reads \cite{28arti3}, \cite{effec}%
\begin{equation}
V_{1}\left( \phi \right) =\sum\limits_{i}\frac{\tilde{n}_{i}}{64\pi ^{2}}%
m_{i}^{4}\left( \phi \right) \left[ \ln \frac{m_{i}^{2}\left( \phi \right) }{%
\mu ^{2}}-C_{i}\right]
\end{equation}%
where $\mu $ is th renormalization scale and $\tilde{n}%
_{i}=n_{i}(-1)^{2S_{i}},$ with $S_{i\text{ }}$and $n_{i\text{ }}$stand for
the spin and the degrees of freedom of the particle $i$. The sum is over all
particles of the model having a mass $m_{i}(\phi )$. It is worthwhile mentionning,
that if we work in the Landau gauge and use the$\overline{\text{ }DR}$
renormalization scheme \cite{renor}, the coefficients $C_{i}$ take the same
value $\frac{3}{2}$ for all kind of particles.
Following ref. \cite{28arti3} \ $\Delta V_{1}^{(T)}\left( \phi,T\right) $
has the form :%
\begin{equation}
\Delta V_{1}^{(T)}\left( \phi,T\right) =\sum_{i=boson}n_{i}\frac{T^{4}}{%
2\pi ^{2}}J_{B}\left( \frac{m_{i}^{2}}{T^{2}}\right) -\sum_{j=fermion}n_{j}%
\frac{T^{4}}{2\pi ^{2}}J_{F}\left( \frac{m_{j}^{2}}{T^{2}}\right)
\end{equation}%
where $J_{B}(x)$, $J_{F}(x)$ are  are respectively the bosonic and fermionic thermal functions,
with the following expansions (for $x\ll 1)$ \cite{29arti3}%
\begin{equation}
J_{B}(x^{2})=-\frac{\pi ^{4}}{45}+\frac{\pi ^{2}}{12}x^{2}-\frac{\pi }{6}%
x^{3}-\frac{1}{32}x^{4}\ln (\frac{x^{2}}{a_{b}})+O(x^{3})  \label{EQB}
\end{equation}%
and%
\begin{equation}
J_{F}(x^{2})=-\frac{7\pi ^{2}}{360}+\frac{\pi ^{2}}{24}x^{2}-\frac{1}{32}%
x^{4}-\frac{1}{32}x^{4}\ln (\frac{x^{2}}{a_{f}})+O(x^{3})  \label{EQF}
\end{equation}%
\ with $\ln (a_{b})$ $\simeq 5.4076$ and $\ln (a_{f})$ $\simeq 2.6351$. For $%
x\gg 1$ one has \cite{29arti3}%
\begin{equation}
J_{B}(x^{2})\simeq J_{F}(x^{2})=\left( \frac{x}{2\pi }\right) ^{\frac{3}{2}%
}e^{-x}\left( 1+\frac{15}{8x}+O\left( x^{-2}\right) \right)
\end{equation}
Regarding the particles content of the compact $341$ model, the effective
potential $V_{eff}(\phi,T)$ can be shown as having the following expression%
\begin{align}
V_{eff}(\phi,T)& =\frac{\mu _{\eta }^{2}\upsilon _{\eta }^{2}+\mu _{\rho
}^{2}\upsilon _{\rho }^{2}+\mu _{\chi }^{2}\upsilon _{\chi }^{2}}{2}+\frac{%
\lambda _{1}\upsilon _{\eta }^{2}+\lambda _{2}\upsilon _{\rho }^{2}+\lambda
_{3}\upsilon _{\chi }^{2}+\lambda _{4}\upsilon _{\eta }^{2}\upsilon _{\rho
}^{2}+\lambda _{5}\upsilon _{\eta }^{2}\upsilon _{\chi }^{2}+\lambda
_{6}\upsilon _{\chi }^{2}\upsilon _{\rho }^{2}}{4} \\
& +\sum\limits_{i=bosons,fermions}\tilde{n}_{i}\frac{m_{i}^{4}}{64\pi ^{2}}%
\left[ \ln \frac{m_{i}^{2}}{\mu ^{2}}-\frac{3}{2}\right] +\frac{T^{4}}{2\pi
^{2}}\sum_{i=bosons,fermions}\tilde{n}_{i}J_{B,F}\left( \frac{m_{i}}{T^{2}}%
\right),  \notag
\end{align}%
with%
\begin{equation}
\tilde{n}_{h}=1,\ \ \ \ \tilde{n}_{ch\arg ed}=2,\text{ \ \ \ }\tilde{n}%
_{quark}=-12,\ \ \ \ \tilde{n}_{_{Z,\text{ }Z^{^{\prime }},\text{ }%
Z^{^{^{\prime \prime }}},\text{ }K,\text{ }K^{^{\prime }}}}=3,\ \ \ \ \tilde{%
n}_{_{W,\text{ }K_{1},\text{ }Y,\text{ }X,\text{ }V}}=6.
\end{equation}

\subsection{Electroweak Phase Transition in the compact $341$ Model}

The electroweak baryogenesis (EWBG) is one of the most attractive and important
 ways of accounting for the observed baryon asymmetry of the universe.
 In order to explain the problem of  baryogenesis, Sakharov posited
three conditions for generating this asymmetry (matter-antimatter) \cite{SAK 1}, 
\cite{depend higgs}. The process causing the violation of the baryon number 
must come from a rapid transition of the sphalons into the symmetrical phase. 
It must violate C and CP since its conservation would lead to the
creation of the same amount of matter-antimatter particles. The involved
interactions should take place out-of-equilibrium so that the last Sakharov
condition is reached by a strong first order phase transition.

Using the high temperature expansions in eqs.(\ref{EQB}), (\ref{EQF}) one
can rewrite eq.(\ref{poteff}) in a simplified form illustrating the
thermal corrections in the effective potential, as in ref.\cite{29arti3}
\begin{equation}
V_{eff}(\phi,T)=D(T^{2}-T_{0}^{2})\phi ^{2}-ET\phi ^{3}+\frac{\lambda }{4}%
\phi ^{4}
\end{equation}%
where $D,$ $E$, $T_{0}$ are temperature independent coeffients and $\lambda $
is a slowly-varying function of $T$. If $E=0$ the phase transition is of a
second order type \cite{Probar} with a transition temperature $T_{0}$ and
the Higgs expectation value $\left\langle \phi \right\rangle _{0}$ such that%
\begin{equation}
\left\langle \phi \right\rangle _{0}=T_{0}\sqrt{\frac{2D}{\lambda }\left( 1-%
\frac{T^{2}}{T_{0}^{2}}\right) }
\end{equation}%
For $E\neq 0$ the phase transition becomes first order, at very high
temperatures $T\gg T_{1}.$ The only minimum of the effective potential is $%
\left\langle \phi \right\rangle _{0}=0.$ When $T=T_{1}$ the effective
potential acquires an extra minimum at a value%
\begin{equation}
\left\langle \phi \right\rangle _{1}=\frac{3ET}{2\lambda }  \label{secondM}
\end{equation}%
which appears as an inflection point at a temperature%
\begin{equation}
T_{1}^{2}=\frac{8\lambda D}{8\lambda D-9E^{2}}T_{0}^{2}
\end{equation}%
As the temperature drops, the second minimum in eq.(\ref{secondM})
becomes deeper, and it degenerates with the other one $\left\langle \phi
\right\rangle _{1}=0.$ The two minima are separated by a potential barrier
at a critical temperature $T_{c}$ such that%
\begin{equation}
T_{c}^{2}=\frac{\lambda DT_{0}^{2}}{\lambda D-E^{2}}
\end{equation}%
The value of the minimum of the critical temperature is given by%
\begin{equation}
\left\langle \phi \right\rangle _{c}=\frac{2ET_{c}}{\lambda },
\end{equation}
The first order phase transition is characterized by the ratio $\frac{%
\left\langle \phi \right\rangle _{c}}{T_{c}}$. The transition from the local
minimum at $\left\langle \phi \right\rangle =0$ to a deeper minimum at $%
\left\langle \phi \right\rangle \neq 0$ proceeds via the thermal tunnelling 
\cite{32arti3}. It can be understood in terms of bubbles formation of the
broken phase in the sea of the symmetrical phase. When enough large bubble
forms and expands until it collides with other bubbles, then the universe
becomes filled with the broken phase. The electroweak phase transition
occurs at a temperature $T_{w}$ $\sim 100$ $GeV$ and must be strongly first
order to achieve a successful EWBG and its quantitative condtion is $\frac{
\phi _{c}}{T_{c}}\geq 1$. The height of the barrier between the two minima
measures of the strength of the transition. Thus, the strong first order phase
transition presents a high barrier.

The EWBG would not be possible at the critical temperature if a second order
phase transition occurred when there was no barrier and the Higgs field would
continuously drop from zero to a non-zero expectation value and that there was
no bubble nucleation which is the source of non-thermal equilibrium. We
would expect a very large baryon violation rate because there is no barrier
between the two vacua, so the sphaleron transition is quick in this case.

In our work, we only consider the contributions of the gauge bosons, top
and exotic quarks, three neutral and three charged Higgses to the
effective potential during the three steps of the spontaneous symmetry breaking.

\subsubsection{The first step phase transition}

The effective potential at finite temperature of the first step of phase
transition ($SU\left( 4\right) _{L}\otimes U\left( 1\right) _{X}$\ $\underset
{\upsilon _{x}}{\rightarrow }SU\left( 3\right) _{L}\otimes U\left( 1\right)
_{X}$) has the compact form
\begin{equation}
V_{eff}(\upsilon _{\chi })=D(T^{2}-T_{0}^{2})\upsilon _{\chi
}^{2}-ET\upsilon _{\chi }^{3}+\frac{\lambda (T)}{4}\upsilon _{\chi }^{4}
\label{Veff1}
\end{equation}
where this phase transition occurs at the TeV scale. The parameters of the
above equation are shown to have the following expressions:
\begin{eqnarray}
D &=&\frac{1}{24\upsilon _{\chi }^{2}}\left( 6m_{V^{\pm \pm
}}^{2}+6m_{Y^{\pm }}^{2}+6m_{X^{\pm }}^{2}+2m_{h_{2}^{\pm }}^{2}+2m_{h^{\pm
\pm }}^{2}\right.  \label{step1} \\
&&\left.
+m_{H_{2}^{0}}^{2}+m_{H_{3}^{0}}^{2}+6m_{J_{1}}^{2}+6m_{J_{2}}^{2}+6m_{J_{3}}^{2}\right)
\notag \\
E &=&\frac{1}{12\pi \upsilon _{\chi }^{3}}\left( 6m_{V^{\pm \pm
}}^{3}+6m_{Y^{\pm }}^{3}+6m_{X^{\pm }}^{3}\right.  \notag \\
&&\left. +2m_{h_{2}^{\pm }}^{3}+2m_{h^{\pm \pm
}}^{3}+m_{H_{2}^{0}}^{3}+m_{H_{3}^{0}}^{3}\right)  \notag \\
T_{0}^{2} &=&\frac{(m_{H_{2}^{0}}^{2}+m_{H_{3}^{0}}^{2})}{4D}-\frac{1}{
32D\pi ^{2}\upsilon _{\chi }^{2}}\left( 6m_{Y^{\pm }}^{4}+6m_{V^{\pm \pm
}}^{4}+6m_{X^{\pm }}^{4}\right.  \notag \\
&&\left.
-12m_{J_{1}}^{4}-12m_{J_{2}}^{4}-12m_{J_{3}}^{4}+m_{H_{2}^{0}}^{4}
+m_{H_{3}^{0}}^{4}+2m_{h_{2}^{\pm }}^{4}+2m_{h^{\pm \pm }}^{4}\right)
\notag
\end{eqnarray}%
\begin{eqnarray*}
\lambda \left( T\right) &=&\frac{\left(
m_{H_{2}^{0}}^{2}+m_{H_{3}^{0}}^{2}\right) }{2\upsilon _{\chi }^{2}}\left\{
1-\left( \frac{1}{8\pi ^{2}\upsilon _{\chi
}^{2}(m_{H_{2}^{0}}^{2}+m_{H_{3}^{0}}^{2})}\right) \left[ 6m_{Y^{\pm
}}^{4}\log \frac{m_{Y^{\pm }}^{2}}{A_{B}T^{2}}\right. \right. \\
&&\left. +6m_{X^{\pm }}^{4}\log \frac{m_{X^{\pm }}^{2}}{A_{B}T^{2}}%
+6m_{V^{\pm \pm }}^{4}\log \frac{m_{V^{\pm \pm }}^{2}}{A_{B}T^{2}}%
-12m_{J_{1}}^{4}\log \frac{m_{J_{1}}^{2}}{A_{F}T^{2}}\right. \\
&&\left. -12m_{J_{2}}^{4}\log \frac{m_{J_{2}}^{2}}{A_{F}T^{2}}%
-12m_{J_{3}}^{4}\log \frac{m_{J_{3}}^{2}}{A_{F}T^{2}}+m_{H_{2}^{0}}^{4}\log 
\frac{m_{H_{2}^{0}}^{2}}{A_{B}T^{2}}\right. \\
&&\left. \left. +m_{H_{3}^{0}}^{4}\log \frac{m_{H_{3}^{0}}^{2}}{A_{B}T^{2}}%
+2m_{h_{2}^{\pm }}^{4}\log \frac{m_{h_{2}^{\pm }}^{2}}{A_{B}T^{2}}%
+2m_{h^{\pm \pm }}^{4}\log \frac{m_{h^{\pm \pm }}^{2}}{A_{B}T^{2}}\right]
\right\}
\end{eqnarray*}%
where the critical temperature $T_{c_{1}}$ is given by%
\begin{equation}
T_{c_{1}}=\frac{T_{0}}{\sqrt{1-E^{2}/\lambda D}}
\end{equation}%
and the condition of first EWPT is 
\begin{equation}
E>\frac{\lambda (T_{c_{1}})}{2}  \label{criti}
\end{equation}

\subsubsection{The second step phase transition}

The effective potential at finite temperature of the second step of phase
transition ($SU\left( 3\right) _{L}\otimes U\left( 1\right) _{X}$\ \ $%
\underset{\upsilon _{_{\eta }}}{\rightarrow }SU\left( 2\right) _{L}\otimes
U\left( 1\right) _{Y}$) reads%
\begin{equation}
V_{eff}(\upsilon _{\eta })=D^{^{\prime }}(T^{2}-T_{0}^{^{^{\prime
}}2})\upsilon _{\eta }^{2}-E^{^{\prime }}T\upsilon _{\eta }^{3}+\frac{%
\lambda ^{^{\prime }}(T)}{4}\upsilon _{\eta }^{4}  \label{Veff2}
\end{equation}%
where this phase transition also occurs  at the TeV scale. The parameters $%
D^{^{\prime }}$, $E^{^{\prime }}$, $T_{0}^{^{^{\prime }}}$ and $\lambda
^{^{\prime }}(T)$ are:%
\begin{eqnarray}
D^{^{\prime }} &=&\frac{1}{24\upsilon _{\eta }^{2}}\left(
3m_{K}^{2}+3m_{K^{^{\prime }}}^{2}+6m_{K_{1}^{\pm }}^{2}+6m_{Y^{\pm
}}^{2}+3m_{Z^{\prime }}^{2}+3m_{Z^{^{\prime \prime }}}^{2}\right.
\label{ste} \\
&&\left. +6m_{U_{1}}^{2}+6m_{D_{2}}^{2}+6m_{D_{3}}^{2}+2m_{h_{1}^{\pm
}}^{2}+2m_{h_{2}^{\pm }}^{2}\right)  \notag \\
E^{^{\prime }} &=&\frac{1}{12\pi \upsilon _{\eta }^{3}}\left( 6m_{Y^{\pm
}}^{3}+6m_{K_{1}^{\pm }}^{3}+3m_{Z^{^{\prime }}}^{3}+3m_{Z^{^{\prime \prime
}}}^{3}\right.  \notag \\
&&\left. +3m_{K}^{3}+3m_{K^{^{\prime }}}^{3}+2m_{h_{1}^{\pm
}}^{3}+2m_{h_{2}^{\pm }}^{3}\right)  \notag \\
T_{0}^{^{^{\prime }}2} &=&-\frac{1}{32\pi ^{2}D^{^{\prime }}\upsilon _{\eta
}^{2}}\left( 6m_{Y^{\pm }}^{4}+6m_{K_{1}^{\pm }}^{4}+3m_{Z^{^{\prime
}}}^{4}+3m_{Z^{^{\prime \prime }}}^{4}+3m_{K}^{4}\right.  \notag \\
&&\left. +3m_{K^{^{\prime
}}}^{4}-12m_{U_{1}}^{4}-12m_{D_{2}}^{4}-12m_{D_{3}}^{4}+2m_{h_{1}^{\pm
}}^{4}+2m_{h_{2}^{\pm }}^{4}\right)  \notag
\end{eqnarray}%
\begin{eqnarray*}
\lambda ^{^{\prime }}(T) &=&\frac{1}{2\upsilon _{\eta }^{2}}\left[ \upsilon
_{\eta }^{2}-\frac{1}{8\pi ^{2}\upsilon _{\eta }^{2}}\left( 6m_{Y^{\pm
}}^{4}\log \frac{m_{Y^{\pm }}^{2}}{A_{B}T^{2}}+6m_{K_{1}^{\pm }}^{4}\log 
\frac{m_{K^{\pm }}^{2}}{A_{B}T^{2}}\right. \right. \\
&&\left. \left. +3m_{Z^{^{\prime }}}^{4}\log \frac{m_{Z^{^{\prime }}}^{2}}{%
A_{B}T^{2}}+3m_{Z^{^{^{\prime \prime }}}}^{4}\log \frac{m_{Z^{^{\prime
\prime }}}^{2}}{A_{B}T^{2}}+3m_{K}^{4}\log \frac{m_{K}^{2}}{A_{B}T^{2}}%
\right. \right. \\
&&\left. \left. +3m_{K^{^{\prime }}}^{4}\log \frac{m_{K^{^{\prime }}}^{2}}{%
A_{B}T^{2}}-12m_{U_{1}}^{4}\log \frac{m_{U_{1}}^{2}}{A_{F}T^{2}}%
-12m_{D_{2}}^{4}\log \frac{m_{D_{2}}^{2}}{A_{F}T^{2}}\right. \right. \\
&&\left. \left. -12m_{D_{3}}^{4}\log \frac{m_{D_{3}}^{2}}{A_{F}T^{2}}%
+2m_{h_{1}^{\pm }}^{4}\log \frac{m_{h_{1}^{\pm }}^{2}}{A_{B}T^{2}}%
+2m_{h_{2}^{\pm }}^{4}\log \frac{m_{h_{2}^{\pm }}^{2}}{A_{B}T^{2}}\right) %
\right]
\end{eqnarray*}%
where the critical temperature $T_{c_{2}}$ is given by
\begin{equation}
T_{c_{2}}=\frac{T_{0}^{^{\prime }}}{\sqrt{1-E^{^{\prime }2}/\lambda
^{^{\prime }}D^{^{\prime }}}}
\end{equation}%
and the condition of first EWPT is%
\begin{equation}
E^{^{\prime }}>\frac{\lambda (T_{c_{2}})}{2}
\end{equation}

\subsubsection{The third step phase transition}

The effective potential at finite temperature of the third step of phase
transition ($SU\left( 2\right) _{L}\otimes U\left( 1\right) _{Y}\ \underset{%
\upsilon _{_{\rho }}}{\rightarrow }U\left( 1\right) _{QED}$) has the form%
\begin{equation}
V_{eff}(\upsilon _{\rho })=D^{^{\prime \prime }}(T^{2}-T_{0}^{^{\prime
\prime }2})\upsilon _{\rho }^{2}-E^{^{\prime \prime }}T\upsilon _{\rho }^{3}+%
\frac{\lambda ^{^{\prime \prime }}(T)}{4}\upsilon _{\rho }^{4}  \label{Veff3}
\end{equation}%
where this phase transition happens at the GeV scale and the coefficients $%
D^{^{\prime \prime }},$ $E^{^{\prime \prime }},$ $T_{0}^{^{^{\prime \prime
}}}$ and $\lambda ^{^{\prime \prime }}(T)$ are:%
\begin{eqnarray}
D^{^{\prime \prime }} &=&\frac{1}{24\upsilon _{\rho }^{2}}\left( 6m_{W^{\pm
}}^{2}+6m_{K_{1}^{\pm }}^{2}+6m_{V^{\pm \pm }}^{2}+3m_{Z}^{2}\right. 
\label{step3} \\
&&\left. +m_{H_{1}^{0}}^{2}+2m_{h_{1}^{\pm }}^{2}+2m_{h^{\pm \pm
}}^{2}+6m_{t}^{2}\right)   \notag \\
E^{^{\prime \prime }} &=&\frac{1}{12\pi \upsilon _{\rho }^{3}}\left(
6m_{W^{\pm }}^{3}+6m_{K_{1}^{\pm }}^{3}+6m_{V^{\pm \pm }}^{3}\right.   \notag
\\
&&\left. +3m_{Z}^{3}+m_{H_{1}^{0}}^{3}+2m_{h_{1}^{\pm }}^{3}+2m_{h^{\pm \pm
}}^{3}\right)   \notag \\
T_{0}^{^{^{\prime \prime }}2} &=&\frac{m_{H_{1}^{0}}^{2}}{4D^{^{\prime
\prime }}}-\frac{1}{32D^{^{\prime \prime }}\pi ^{2}\upsilon _{\rho }^{2}}%
\left( 6m_{W^{\pm }}^{4}+6m_{K_{1}^{\pm }}^{4}+6m_{V^{\pm \pm }}^{4}\right. 
\notag \\
&&\left. +3m_{Z}^{4}+m_{H_{1}^{0}}^{4}+2m_{h_{1}^{\pm }}^{4}+2m_{h^{\pm \pm
}}^{4}-12m_{t}^{4}\right)   \notag
\end{eqnarray}%
\begin{eqnarray*}
\lambda ^{^{\prime \prime }}(T) &=&\frac{m_{H_{1}^{0}}^{2}}{2\upsilon _{\rho
}^{2}}-\frac{1}{16\pi ^{2}\upsilon _{\rho }^{4}}\left( 6m_{W^{\pm }}^{4}\log 
\frac{m_{W^{\pm }}^{2}}{A_{B}T^{2}}+6m_{K_{1}^{\pm }}^{4}\log \frac{%
m_{K_{1}^{\pm }}^{2}}{A_{B}T^{2}}\right.  \\
&&\left. +6m_{V^{\pm \pm }}^{4}\log \frac{m_{V^{\pm \pm }}^{2}}{A_{B}T^{2}}%
+3m_{Z}^{4}\log \frac{m_{Z}^{2}}{A_{B}T^{2}}+m_{H_{1}^{0}}^{4}\log \frac{%
m_{H_{1}^{0}}^{2}}{A_{B}T^{2}}\right.  \\
&&\left. +2m_{h_{1}^{\pm }}^{4}\log \frac{m_{h_{1}^{\pm }}^{2}}{A_{B}T^{2}}%
+2m_{h^{\pm \pm }}^{4}\log \frac{m_{h^{\pm \pm }}^{2}}{A_{B}T^{2}}%
-12m_{t}^{4}\log \frac{m_{t}^{2}}{A_{F}T^{2}}\right) 
\end{eqnarray*}%
with the critical temperature $T_{c_{3}}$ is given by%
\begin{equation}
T_{c_{3}}=\frac{T_{0}^{^{\prime \prime }}}{\sqrt{1-E^{^{\prime \prime
}2}/\lambda ^{^{\prime \prime }}D^{^{\prime \prime }}}}
\end{equation}%
and the condition of the third step of phase transition is 
\begin{equation}
E^{^{\prime \prime }}>\frac{\lambda (T_{c_{3}})}{2}
\end{equation}

\section{Numerical Results}
\label{sect4}

Before proceeding, we must first diagonalize first the scalar potential $V(\eta,\rho,\chi )$. To do so, 
we must find the eingenvalues of the block
matrix $M_{6\times 6}$ written in the basis $\Phi _{1}=\eta ^{+}\eta $, $%
\Phi _{2}=\rho ^{+}\rho $, $\Phi _{3}=\chi ^{+}\chi $, $\Phi _{4}=\rho
^{+}\eta $, $\Phi _{5}=\chi ^{+}\eta $,$\Phi _{6}=\chi ^{+}\rho $:%
\begin{equation}
M_{6\times 6}=\left( 
\begin{array}{cc}
\left( 
\begin{array}{ccc}
\lambda _{1} & \lambda _{4}/2 & \lambda _{5}/2 \\ 
\lambda _{4}/2 & \lambda _{2} & \lambda _{6}/2 \\ 
\lambda _{5}/2 & \lambda _{6}/2 & \lambda _{3}%
\end{array}%
\right) & 0_{3\times 3} \\ 
0_{3\times 3} & \left( 
\begin{array}{ccc}
\lambda _{7} & 0 & 0 \\ 
0 & \lambda _{8} & 0 \\ 
0 & 0 & \lambda _{9}%
\end{array}%
\right)%
\end{array}%
\right)  \label{matrix}
\end{equation}

Straightforward but tedious calculations give the following eigenvalues%
\begin{eqnarray}
V_{1} &=&\pm \lambda _{9}/2,\text{ }V_{2}=\pm \lambda _{8}/2,\text{ }%
V_{3}=\pm \lambda _{7}/2 \\
V_{4} &=&\left\{ 
\begin{array}{c}
-\left( \frac{2}{27}\beta ^{3}-\frac{\gamma \beta }{3}+\delta \right) /2 \\ 
-\left[ \left( 
\begin{array}{c}
4(\gamma -\beta ^{3}/3)^{3} \\ 
+27(\frac{2}{27}\beta ^{3}-\frac{\gamma \beta }{3}+\delta )^{2}%
\end{array}%
\right) /27\right] ^{1/2}/2%
\end{array}%
\right\} ^{1/3}+  \notag \\
&&\left\{ 
\begin{array}{c}
-\left( \frac{2}{27}\beta ^{3}-\frac{\gamma \beta }{3}+\delta \right) /2 \\ 
+\left[ \left( 
\begin{array}{c}
4(\gamma -\beta ^{3}/3)^{3} \\ 
+27(\frac{2}{27}\beta ^{3}-\frac{\gamma \beta }{3}+\delta )^{2}%
\end{array}%
\right) /27\right] ^{1/2}/2%
\end{array}%
\right\} ^{1/3}  \notag \\
V_{5} &=&(-V_{4}+\sqrt{\Delta })/2\text{, }V_{6}=(-V_{4}-\sqrt{\Delta })/2 
\notag
\end{eqnarray}%
where%
\begin{align}
\beta & =-(\lambda _{1}+\lambda _{2}+\lambda _{3}) \\
\gamma & =\left( (\lambda _{6}^{2}+\lambda _{5}^{2}+\lambda
_{4}^{2})/4+\lambda _{1}\lambda _{2}+\lambda _{1}\lambda _{3}+\lambda
_{2}\lambda _{3}\right)  \notag \\
\delta & =\left( \left( \lambda _{6}^{2}\lambda _{1}+\lambda _{5}^{2}\lambda
_{2}+\lambda _{4}^{2}\lambda _{3}\right) /4-\lambda _{1}\lambda _{2}\lambda
_{3}-(\lambda _{4}\lambda _{5}\lambda _{6})/3\right)  \notag
\end{align}%
Now, in order to make our study clear, viable and self-consistent, some
theoretical constraints have to be imposed on the scalar potential
dimensionless couplings $V_{i}(i=\overline{1,6}).$

1)- $V_{i}$ 's have to be real such that the masses of all particles are
real and positive (no ghost masses in the model).

2)- The potential has to be bounded from below and this will impose the
stability conditions $V_{i}> 0$ $(i=\overline{1,6}).$

3)- In order to preserve the perturbative unitarity, $V_{i}$ 's have to
verify the constraints $0<V_{i}<8\pi $ $(i=\overline{1,6}).$

In what follows, we identify the lightest scalar as neutral Higgs like boson
with mass $m_{H_{1}^{0}}=125.616$ GeV, and take the top quark mass $m_{t}=172.4466$
GeV$,$ the $W^{\pm }$, $Z$ gauge bosons masses $m_{W^{\pm }}=80.40$ GeV$,$ $%
m_{Z}=91.6838$ GeV, the VeV $\upsilon _{\rho }=246$ GeV and the gauge
coupling of the standard model $g_{L}\simeq g_{_{X}}\simeq 0.65$. Taking the
Yukawa coupling randomly and using a Monte-Carlo simulation after imposing
the constraint of a strong first order phase transition, we obtain%
\begin{eqnarray}
&&\left. \lambda _{33}^{u}=0.9913\text{ },\text{ }0.385\leq \lambda
_{11}^{J}\leq 0.54\text{\, }0.382\leq \lambda _{22}^{J}\leq 0.536\text{ }
,0.386\text{ }\leq \lambda _{33}^{J}\leq 0.542\right. \\
&&\left. 0.42\leq \lambda _{11}^{U}\leq 0.61\text{, }0.4\leq \lambda
_{22}^{D}\leq 0.54\text{, }0.41\leq \lambda _{33}^{D}\leq 0.56\right.\notag
\end{eqnarray}
Moreover, using the fact that the VeVs take the values $\upsilon _{\chi }\sim \upsilon
_{\eta }\sim 2\ $TeV, (in order to avoid the Landau pole \cite{pol1}, \cite
{pol2}, \cite{pol3}), and the masses of the exotic quarks $U,$ $J_{1},$ $J_{2},\
J_{3},$ $D_{2},\ D_{3}$ are in the range of $600-700$ GeV which are
compatible with the LHC results giving the lower band reproducing the
experimental results (bands from Monte-Carlo simulation is displayed in
table 3), and  imposing the theoretical constraints mentioned previously, we get
the following stangent conditions on the scalar potential dimensionless
coupling constants:
\begin{eqnarray}
&&\left. (\lambda _{2}+\frac{[\lambda _{3}\lambda _{4}^{2}+\lambda
_{6}(\lambda _{1}\lambda _{6}-\lambda _{4}\lambda _{5})]}{\lambda
_{5}^{2}-4\lambda _{1}\lambda _{3}})=0.258\right. \\
&&\left. 0.15\leq \left( \lambda _{1}+\lambda _{3}-\sqrt{(\lambda
_{1}-\lambda _{3})^{2}+\lambda _{5}^{2}}\right) \leq 0.38\right.  \notag \\
&&\left. 0.34\leq \left( \lambda _{1}+\lambda _{3}+\sqrt{(\lambda
_{1}-\lambda _{3})^{2}+\lambda _{5}^{2}}\right) \leq 0.86\right.  \notag \\
&&\left. 0.24\leq (\lambda _{1}+\lambda _{3})\leq 0.62\text{, }0.009\leq
((\lambda _{1}-\lambda _{3})^{2}+\lambda _{5}^{2})\leq 0.057\right.  \notag
\\
&&\left. 0.156\leq \lambda _{7}\leq 0.31\text{, }0.16\leq \lambda _{8}\leq
0.27\text{, }0.167\leq \lambda _{9}\leq 0.32\right.  \notag
\end{eqnarray}%
The later constraints lead to the following intervals on the new heavy Higgses
and gauge bosons masses shown in the table 3.

\begin{figure}[!h]
\begin{center}
\includegraphics[scale=0.5]{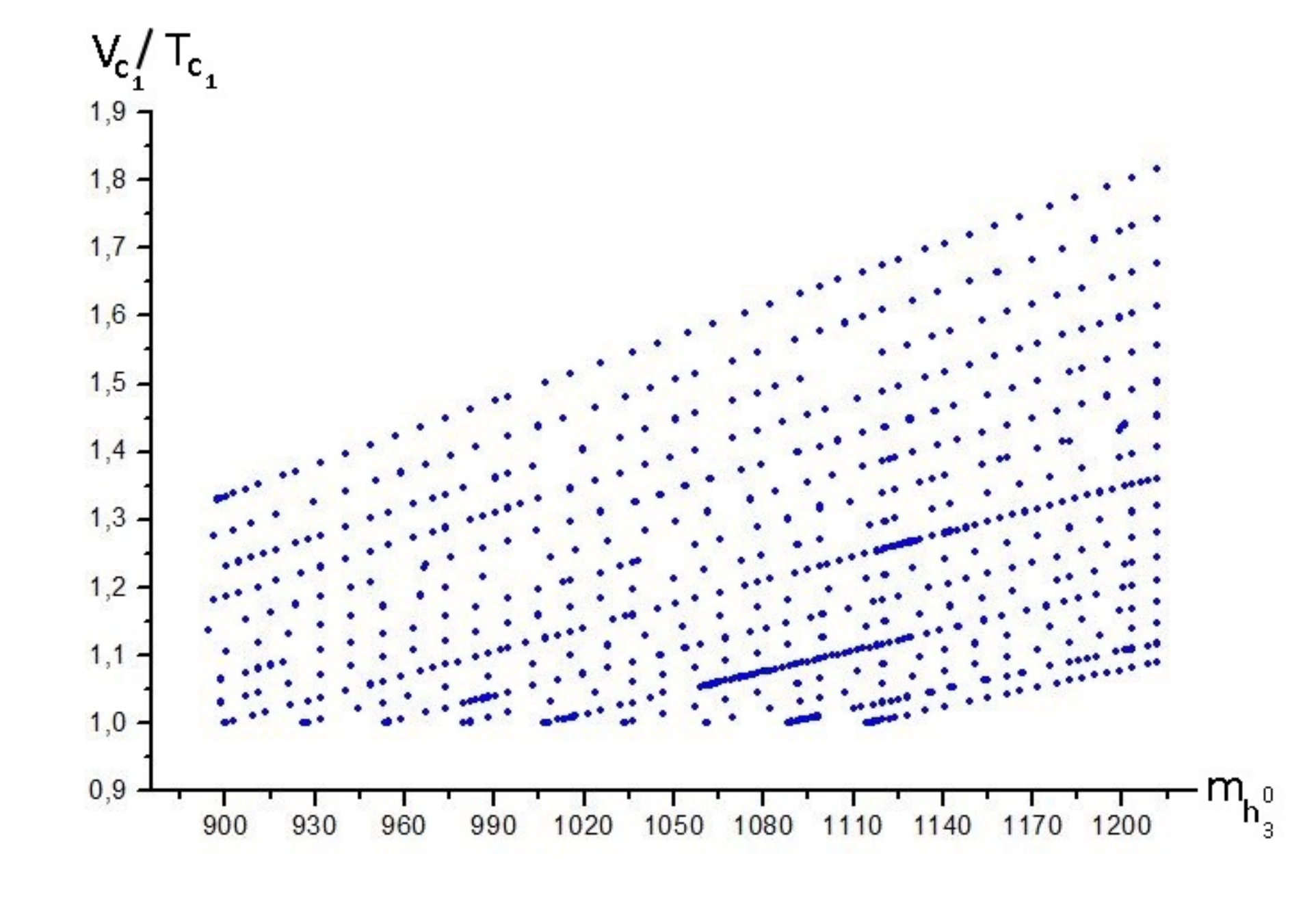} 
\end{center}
\caption{(color online only) The ratio $\frac{\protect\upsilon _{c_{1}}}{T_{c_{1}}}$ in terms of $
m_{h_{3}^{0}}$ for the allowed strong first order EWPT region (density plot).
}
\label{fig1}
\end{figure}

\begin{figure}[!h]
\begin{center}
\includegraphics[scale=0.5]{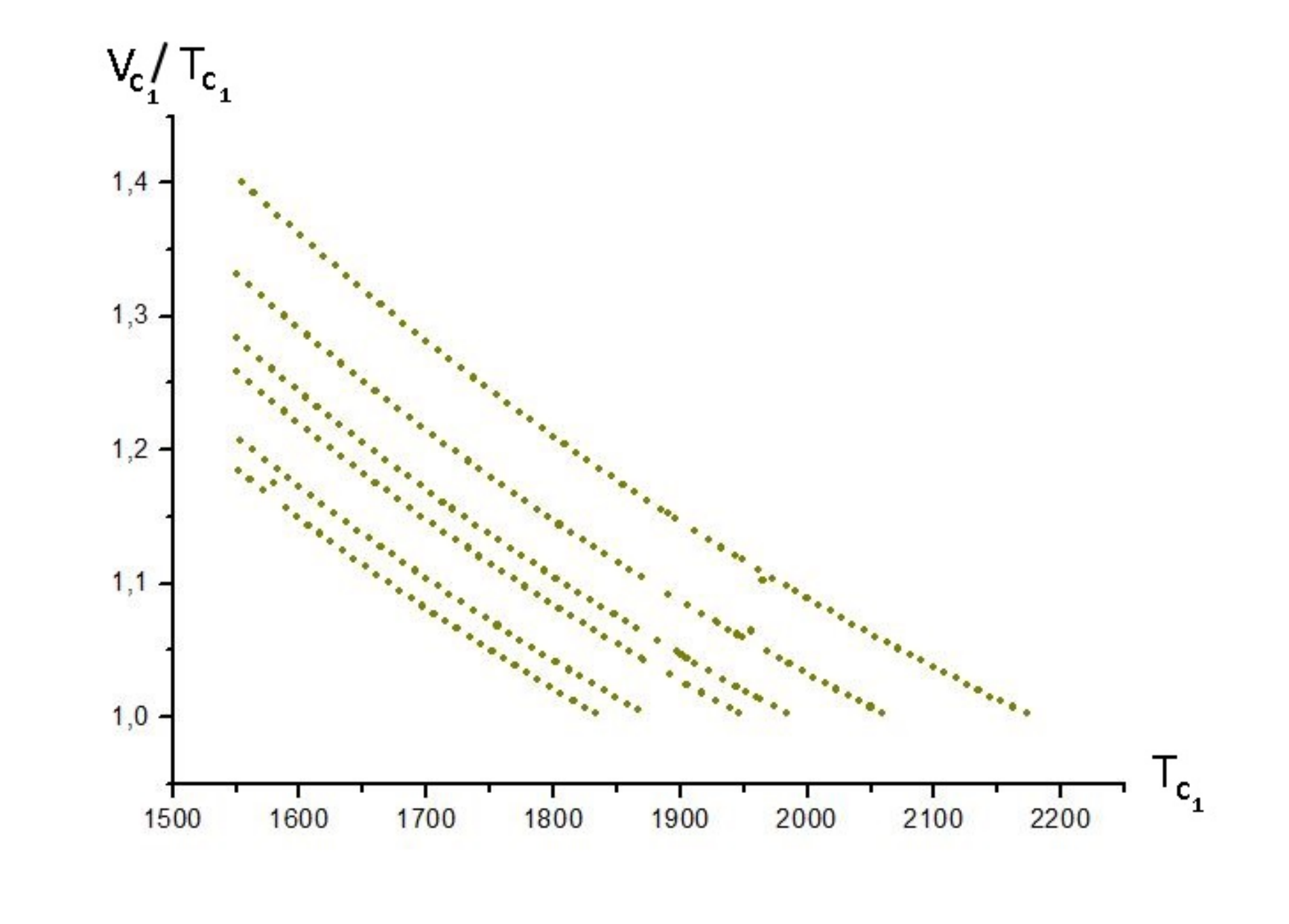} 
\end{center}
\caption{(color online only) The ratio $\frac{\protect\upsilon _{c_{1}}}{T_{c_{1}}}$ in terms of 
$T_{{c}_1}$ for the allowed strong first order EWPT region (density plot).
}
\label{fig2}
\end{figure}

\begin{figure}[!h]
\begin{center}
\includegraphics[scale=0.5]{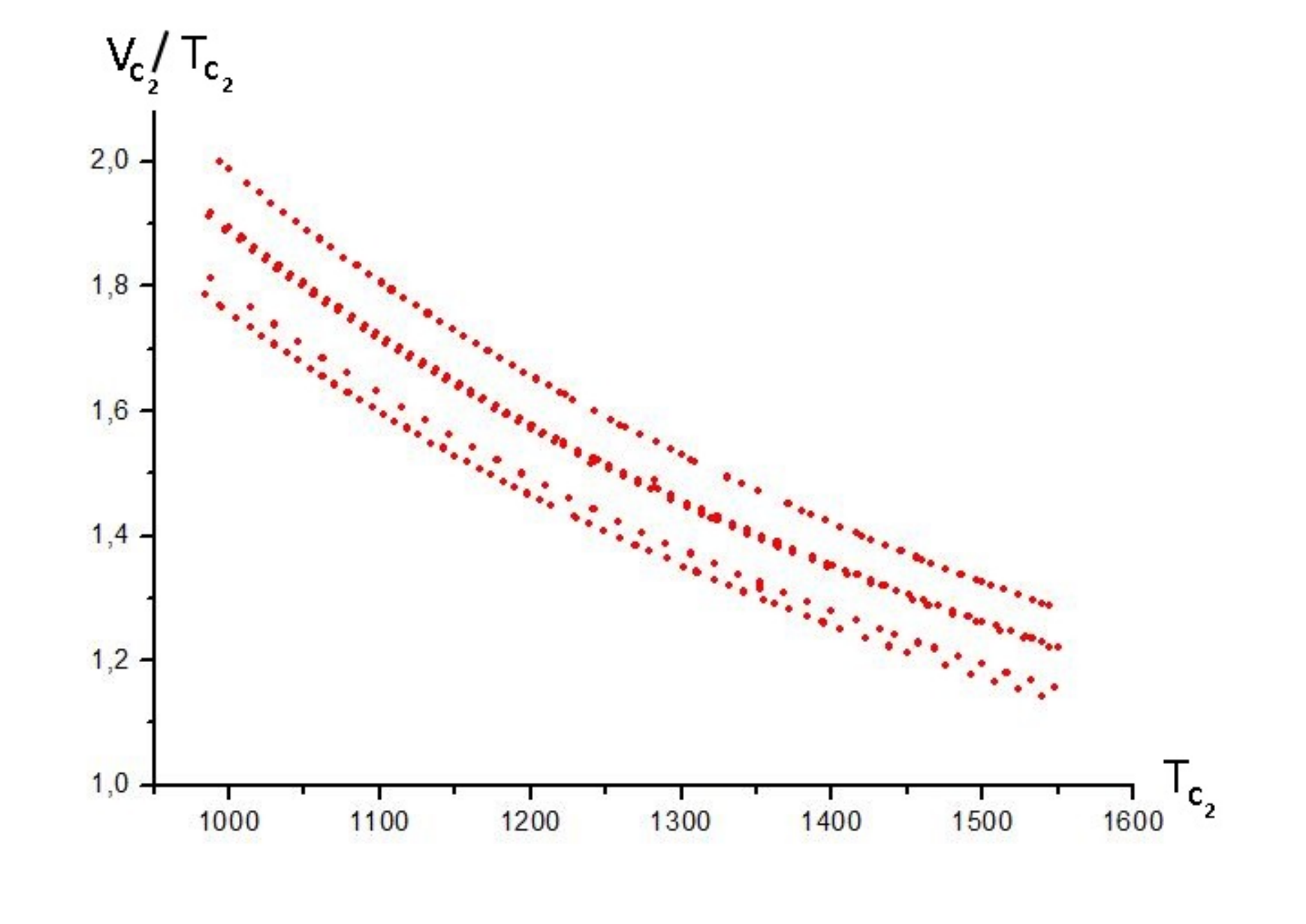} 
\end{center}
\caption{(color online only) The ratio $\frac{\protect
\upsilon _{c_{2}}}{T_{c_{2}}}$ in terms of  $T_{c_{2}}$ for the allowed strong first order EWPT region
(density plot).
}
\label{fig3}
\end{figure}

\begin{figure}[!h]
\begin{center}
\includegraphics[scale=0.5]{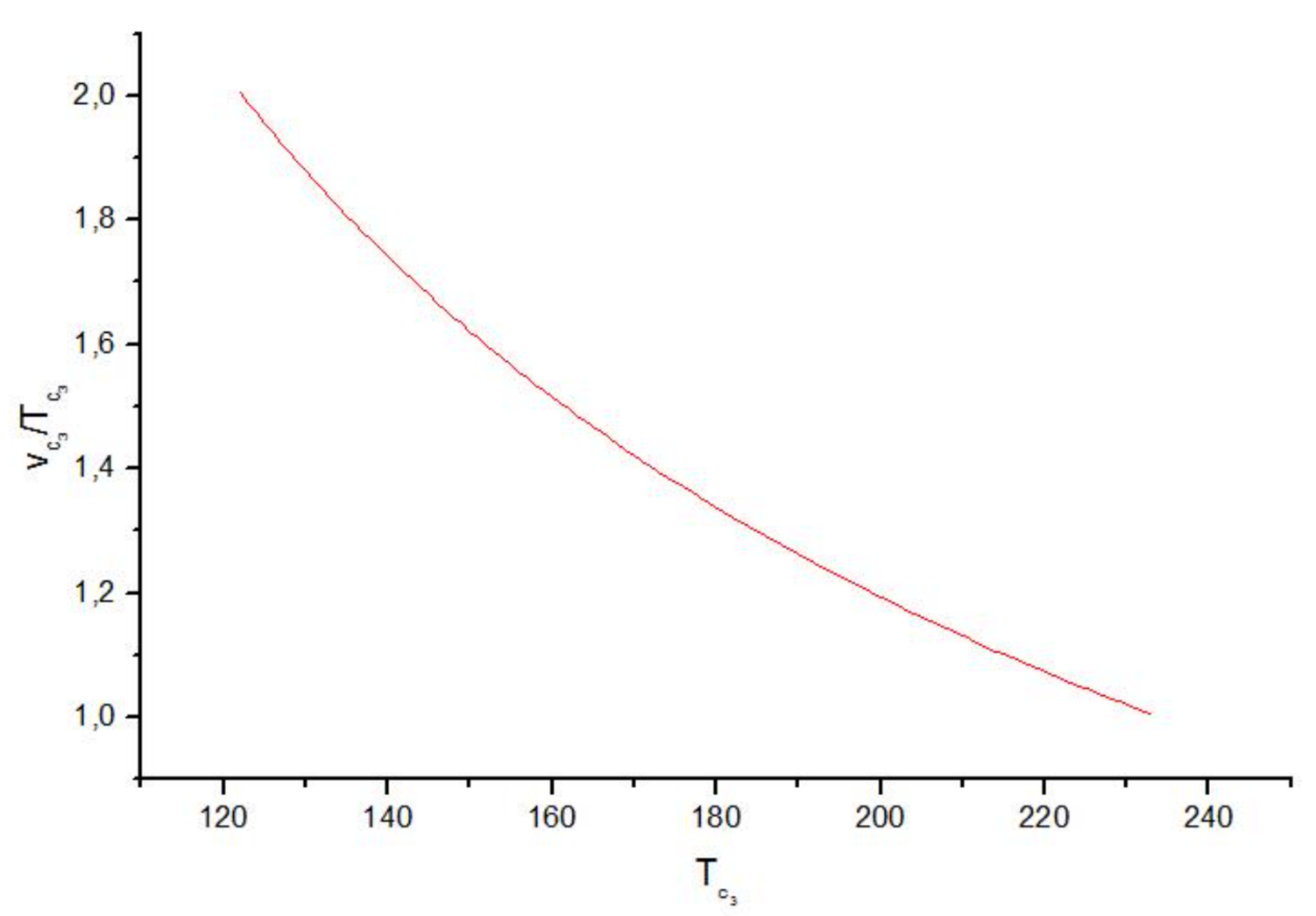} 
\end{center}
\caption{(color online only) The ratio $\frac{\protect\upsilon _{c_{3}}%
}{T_{c_{3}}}$ in terms of  $T_{c_{3}}$
}
\label{fig4}
\end{figure}

Figure \ref{fig1} represents the allowed region of the first step of the
EWPT for the ratio $R_{1}=\upsilon _{c_{1}}/T_{c_{1}}$ as a function of the
heavy neutral Higgs boson mass $m_{h_{3}^{0}},$ where the critical
temperature $T_{c_{1}}$ takes values in the interval $T_{c_{1}}\sim 1200-2000$
GeV. It is worth mentionning that the confidence band (the density plot)
comes from the fact that at a given value of $m_{h_{3}^{0}}$ we still have
many choices of the scalar couplings $\lambda _{i}$'s combined to give the
same value of the ratio $R_{1}$. Figure \ref{fig2} displays the variation of
the ratio $R_{1}$ within the allowed region of the parameters space in
terms of the critical temperature $T_{c_{1}}$ fulfilling the strong first
order EWPT. Figure \ref{fig3} is similar to figure \ref{fig2} but for the
second step of EWPT, where $T_{c_{2}}\sim 989-1549$ GeV. 
Notice that in this case the allowed region is narrower than the one of the first step, this
is due to the fact that the parameters space is more constrained. Figure \ref
{fig4} is similar to figure \ref{fig2} but for the third step of the EWPT, where $T_{c_{3}}\sim 120-235$ GeV.

\begin{figure}[!h]
\begin{center}
\includegraphics[scale=0.5]{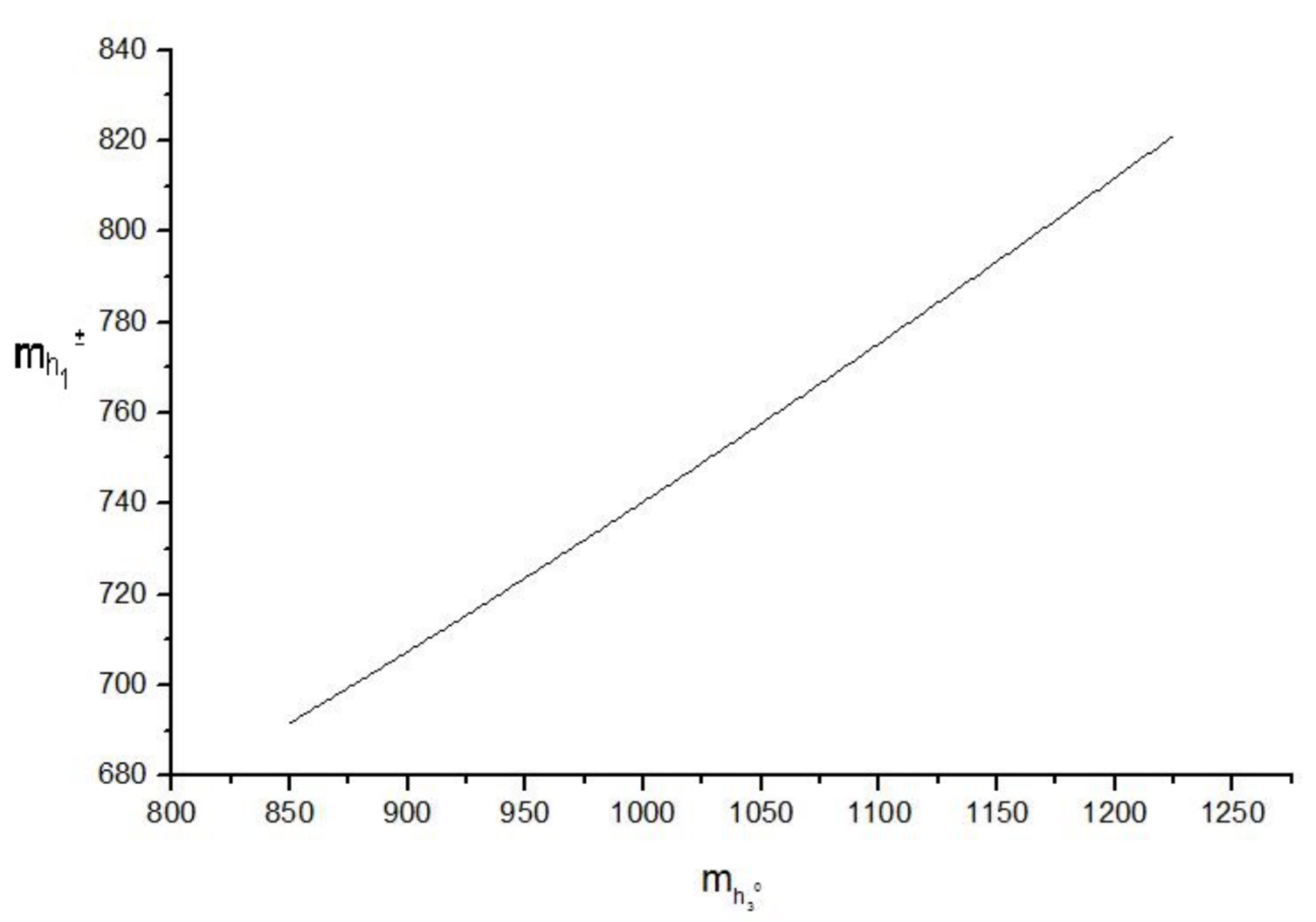} 
\end{center}
\caption{{(color online only) Variation of $m_{h_{1}^{\pm }}$
as a function of $m_{h_{3}^{0}}$ verifying EWPT and theoretical constraints}
}
\label{fig5}
\end{figure}

\begin{figure}[!h]
\begin{center}
\includegraphics[scale=0.5]{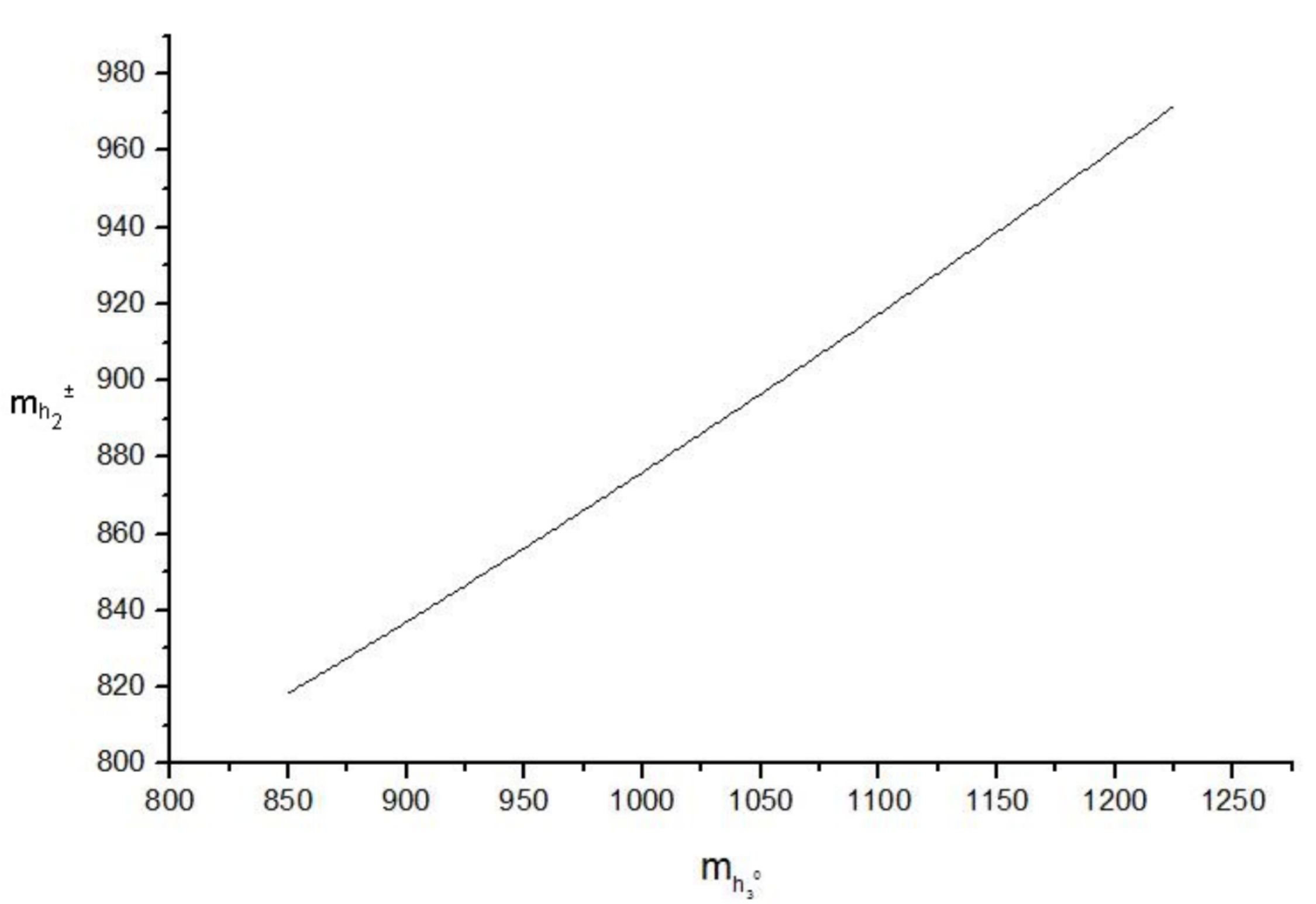} 
\end{center}
\caption{{(color online only) Variation of $m_{h_{2}^{\pm }}$
as a function of $m_{h_{3}^{0}}$ }
}
\label{fig6}
\end{figure}

\begin{figure}[!h]
\begin{center}
\includegraphics[scale=0.5]{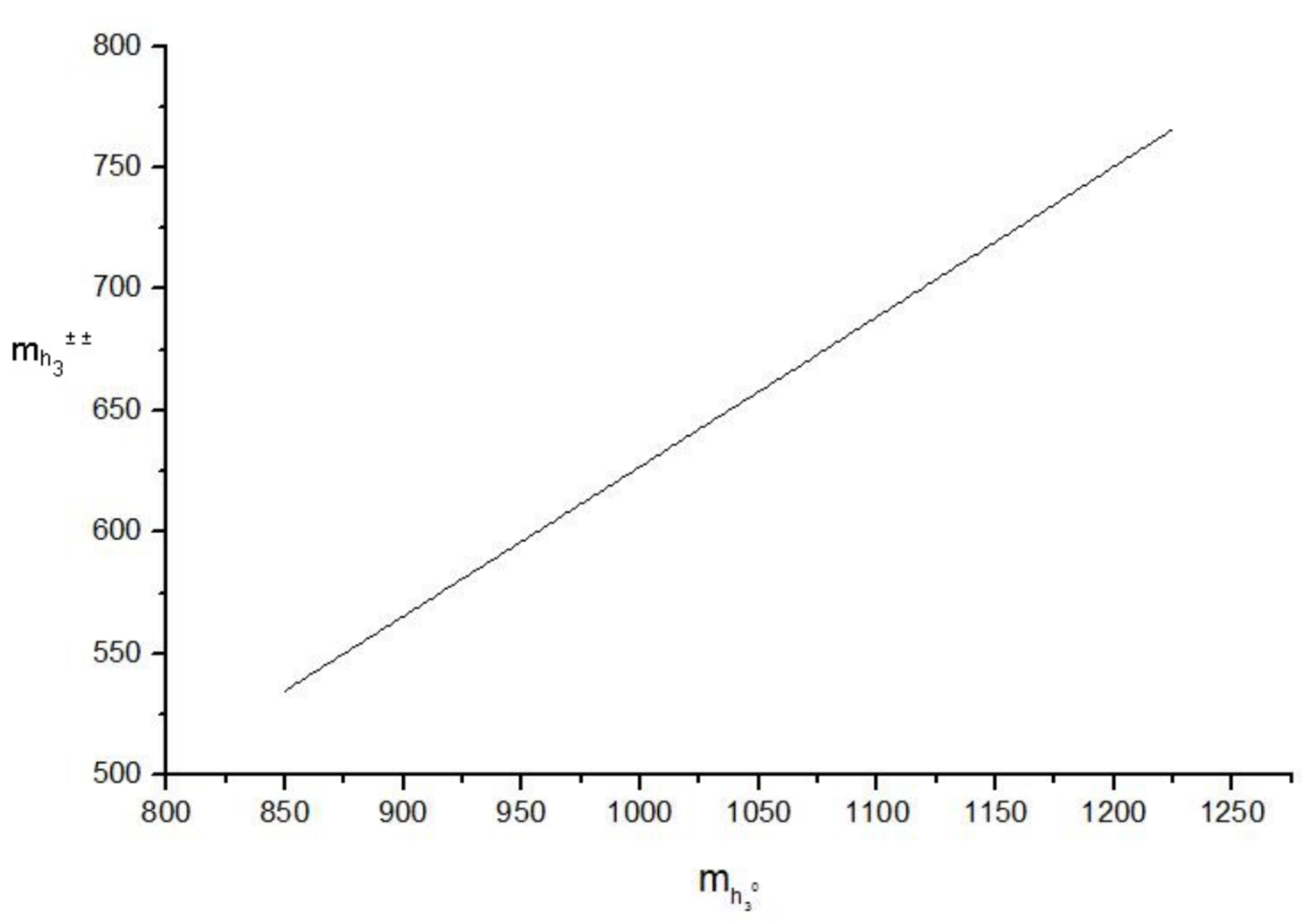} 
\end{center}
\caption{{(color online only) Variation of $m_{h_{3}^{\pm\pm }}$
as a function of $m_{h_{3}^{0}}$ }
}
\label{fig7}
\end{figure}

For the sake of illustration and to get an
idea if we take  $\lambda _{1}\approx 0.2538,$ $\lambda _{2}\approx 0.2809,$ $\lambda _{3}\approx 0.1882,$
$\lambda _{4} \approx 0.1063,$ $\lambda _{5} \approx 0.155$, $\lambda _{6} \approx0.155,$ $\lambda
_{7} \approx 0.2$, $\lambda _{8} \approx 0.2128$, $\lambda _{9} \approx 0.2339,$ $\upsilon _{\eta
}\approx 1987.72$ GeV, $\upsilon _{\chi }\approx1988.915$ GeV, we obtain $%
m_{h_{3}^{0}}\approx 1098.8153$ GeV, $m_{h_{2}^{0}}\approx 735.9784$ GeV, $T_{c_{1}}\approx 1820$
GeV, $T_{c_{2}} \approx 1000$ GeV$,$ $T_{c_{3}} \approx 122$ GeV and the ratio $\upsilon
_{c_{1}}/T_{c_{1}}\approx 1.09283,$ $\upsilon _{c_{2}}/T_{c_{2}}\approx1.98776,$ $%
\upsilon _{c_{3}}/T_{c_{3}}\approx 2.00568$.
Figures \ref{fig5}, \ref{fig6} and  \ref{fig7} represent the variation of the heavy charged and double charged Higgs masses ($m_{h_{1}^{\pm
}},m_{h_{2}^{\pm }}$), and $m_{h^{\pm \pm }}$  respectively of the
model as a function of the heavy neutral Higgs mass $m_{h_{3}^{0}}$
fulfolling the theoretical constraints and the strong first order EWPT.


\section{Sphaleron Rate in the Compact $341$ Model}
\label{sect5}

As it is well established in the literature \cite{Sphaleron331}, \cite{Sphaleron 1},  \cite{Sphaleron 2},  
\cite{Sphaleron 3} and in order to describe correctly the EWB one has to know : 
\begin{enumerate}[label=(\roman*)]
\item the type of the phase transition (first, second order...); 
\item the critical and bubbles nucleation temperatures (giving informations about the occurrence of phase transitions);
\item the sphaleron rate necessary to generate the baryon number asymmetry. 
\end{enumerate}

\begin{center}
\small{Table 3: Various bands of the different particle masses compatible
with strong EWPT and theoretical constraints.}\\
$
\begin{tabular}{|c|c|c|}
\hline
particle & $m_{\min }$(GeV) & $m_{\max }$(GeV) \\ \hline
$m_{h_{1}^{\pm }}$ & $561.29$ & $700$ \\ \hline
$m_{h_{2}^{\pm }}$ & $834.4$ & $939.76$ \\ \hline
$m_{h_{3}^{\pm \pm }}$ & $634.064$ & $749.46$ \\ \hline
$m_{H_{2}^{0}}$ & $600$ & $805$ \\ \hline
$m_{H_{3}^{0}}$ & $896.36$ & $1211.52$ \\ \hline
$m_{U}$ & $590.85$ & $760.11$ \\ \hline
$m_{J_{1}}$ & $592.96$ & $702.67$ \\ \hline
$m_{J_{2}}$ & $588.685$ & $697.59$ \\ \hline
$m_{J_{3}}$ & $594.58$ & $704.58$ \\ \hline
$m_{D_{2}}$ & $570.5$ & $683.55$ \\ \hline
$m_{D_{3}}$ & $585.5$ & $705$ \\ \hline
$m_{k_{1}^{\pm }}$ & $576.79$ & $651.67$ \\ \hline
$m_{k},m_{k^{^{\prime }}}$ & $571.22$ & $646.75$ \\ \hline
$m_{V^{\pm \pm }}$ & $602.61$ & $712.28$ \\ \hline
$m_{Y^{\pm }}$ & $826.46$ & $958.77$ \\ \hline
$m_{X^{\pm }}$ & $597.28$ & $707.78$ \\ \hline
$m_{Z^{^{\prime }}}$ & $695.11$ & $787.01$ \\ \hline
$m_{Z^{"}}$ & $1455.27$ & $1647.67$
 \\ \hline
\end{tabular}
$
\end{center}

In the previous sections, we have discussed extensively the phase transition in the context of the compact $341$ model. 
Furthermore, and as it was mentioned before, sphalerons are one of the most important ingredients in the study of EWB because 
the spaleron rate controls the rate of the baryon number density in the early universe. It should be noted that, the sphaleron phenomenon occurs if 
one has the transition from the zero VeV crossing the barrier to a non-zero VeV without tunnelling (classically).

Following ref.\cite{Sphaleron331}, the sphaleron rate $\Gamma$ by unit time is related to the sphaleron energy $\varepsilon$ via the relation
\begin{equation}
\frac{\Gamma }{V}=\alpha ^{4}T^{4}\exp (-\varepsilon /T)  \label{tau}
\end{equation}
with $T$ is the temperature, $\varepsilon $ is the sphaleron energy, $\alpha=1/30$ 
is a constant and $V$ is the volume of the EWPT region,  $V=\frac{4\pi r^{3}}{3}\sim 
\frac{1}{T^{3}}$. 
To calculate the energy $\varepsilon$, we start from the gauge-Higgs Lagrangian density of the compact $341$ model 
\begin{equation}
\mathcal{L}_{gauge\text{ }-Higgs}=-\frac{1}{4}F_{\mu \nu }^{a}F^{a\mu \nu
}+(D_{\mu }\chi )^{+}(D^{\mu }\chi )+(D_{\mu }\eta )^{+}(D^{\mu }\eta
)+(D_{\mu }\rho )^{+}(D^{\mu }\rho )-V(\chi,\eta,\rho )  \label{lagra}
\end{equation}%
then we derive the Hamiltonian density and deduce the energy $\varepsilon$ to get at the end the following form \cite{Sphaleron331}
\begin{equation}
\varepsilon ={\textstyle\int }d^{3}x\left[ (D_{\mu }\chi )^{+}(D^{\mu }\chi
)+(D_{\mu }\eta )^{+}(D^{\mu }\eta )+(D_{\mu }\rho )^{+}(D^{\mu }\rho
)+V(\chi,\eta,\rho )\right]  \label{Energ}
\end{equation}
using the effective potential analysis at finite temperature with VeV $v_\chi$, $v_\eta$ and $v_\rho$  discussed in section 
\ref{EPT}, eq. (\ref{Energ}) can be rewritten as 
\begin{equation}
\varepsilon =4\pi {\textstyle\int }d^{3}x\left[ \frac{1}{2}(\nabla \upsilon
_{\chi })^{2}+\frac{1}{2}(\nabla \upsilon _{\eta })^{2}+\frac{1}{2}(\nabla
\upsilon _{\rho })^{2}+V_{eff}(\chi,\eta,\rho )\right]  \label{aqq}
\end{equation}
using the static field approximation \cite{Sphaleron331}

\begin{equation}
\frac{\partial \upsilon _{\chi }}{\partial t}=\frac{\partial \upsilon _{\eta
}}{\partial t}=\frac{\partial \upsilon _{\rho }}{\partial t}=0  \label{appro}
\end{equation}%
together with the equations of motion of the VeVs $v_\chi$, $v_\eta$ and $v_\rho$, leads to the following expressions of the 
sphaleron energy $\varepsilon$ for each step of the phase transition : 

\begin{equation}
\varepsilon _{sph\text{ }(SU(4),(SU(3),SU(2))} =4\pi {\textstyle\int }\left[ \frac{1}{2}
\frac{d^{2}\upsilon _{\chi, \eta, \rho }}{dr^{2}}+V_{eff}(\upsilon _{\chi, \eta, \rho },T)\right]
r^{2}dr  \label{eq4}
\end{equation}


To estimate the sphaleron rate $\Gamma$, and as it was done in ref. \cite{Sphaleron331} concerning the $331$ model, we proceed with the 
following approximation :
\begin{enumerate}
\item{Static approximation}:  We assume that the VeVs of Higgs fields do not change from point to point of the universe, that is $
{\vec\nabla v_{\chi, \eta, \rho}}=0$ corresponding to the extremal of $V_{eff}$ when using the field equations. The later are 
shown to be reduced to  
\begin{equation}
\frac{\partial V_{eff}(\upsilon _{\chi, \eta, \rho })}{\partial \upsilon _{\chi \eta, \rho  }}=0.
\end{equation}
Thus, the sphaleron energies of eq.(\ref{eq4}) are simplified and  obtain
\begin{equation}
\varepsilon _{sph\text{ }(SU(4),(SU(3),SU(2))} =\frac{4 \pi r^3}{3} V_{eff}(v_{\chi, \eta, \rho}, T).
\end{equation}
Using eqs. (\ref{Veff1}), (\ref{Veff2}) and (\ref{Veff3}) together with  ${\vec\nabla v_{\chi, \eta, \rho}}=0$, we get the 
following expressions for the sphaleron rates
\begin{equation}
\Gamma= \left\{
\begin{array}{rcl}
\alpha^4T \exp(-\frac{E^4}{4\lambda^3_T}) & \,\,\,\,\,\,\,\,\,\text{for} \, \, \, SU(4)\rightarrow SU(3)
\\
\alpha^4T \exp(-\frac{E'^4}{4\lambda'^3_T}) & \,\,\,\,\,\,\,\,\,\text{for} \, \, \, SU(3)\rightarrow SU(2)
\\
\alpha^4T \exp(-\frac{E''^4}{4\lambda''^3_T}) & \,\,\,\,\,\,\,\,\,\text{for} \, \, \, SU(2)\rightarrow U(1)
\end{array}
\right.
\end{equation}
It is worth mentionning that, for the heavy particles of the $341$ model within the allowed regions where the strong first order phase 
transitions occur, the quantities $E$ (resp. $E'$,$E''$) and  $\lambda$ (resp. $\lambda'$,$\lambda''$) are almost constant. 
Thus, in this approximation $\Gamma$ becomes linear function of $T$ as illustrated in fig. \ref{fig8}.  Moreover, 
numerical results show that for temperatures below that of the phase transition $T_c$ where the universe switches to the 
symmetry breaking phase, the sphaleron rate is still much larger than the Hubble parameter and this leads to the whashout of the 
$B$-violation.  Fig. \ref{fig8} displays the sphaleron rate as a function of the temperature for the three steps of the SSB. 
Notice that for the first step, if we take $T=1700$ GeV $<T_{c_1}\sim 1800$ GeV, $\Gamma\sim 2.0985 \times 10^{-3}$ $>>H$, and for the 
second step, if we take $T=900$ GeV$<T_{c_2}\sim 1000$ GeV, $\Gamma\sim 1.1401\times 10^{-3}$ $>>H$.  Likewise for the third step, if we 
take  $T=70$ GeV$<T_{c_3}\sim 122$ GeV, $\Gamma\sim 8.654\times 10^{-4}$ $>>H$. Consequently, in this approximation the sphaleron decoupling 
condition cannot be satisfied (same result was obtiened by the authors of ref.\cite{Sphaleron331} in the case of the $331$ model).
\begin{figure}[!h]
\begin{center}
\includegraphics[scale=0.5]{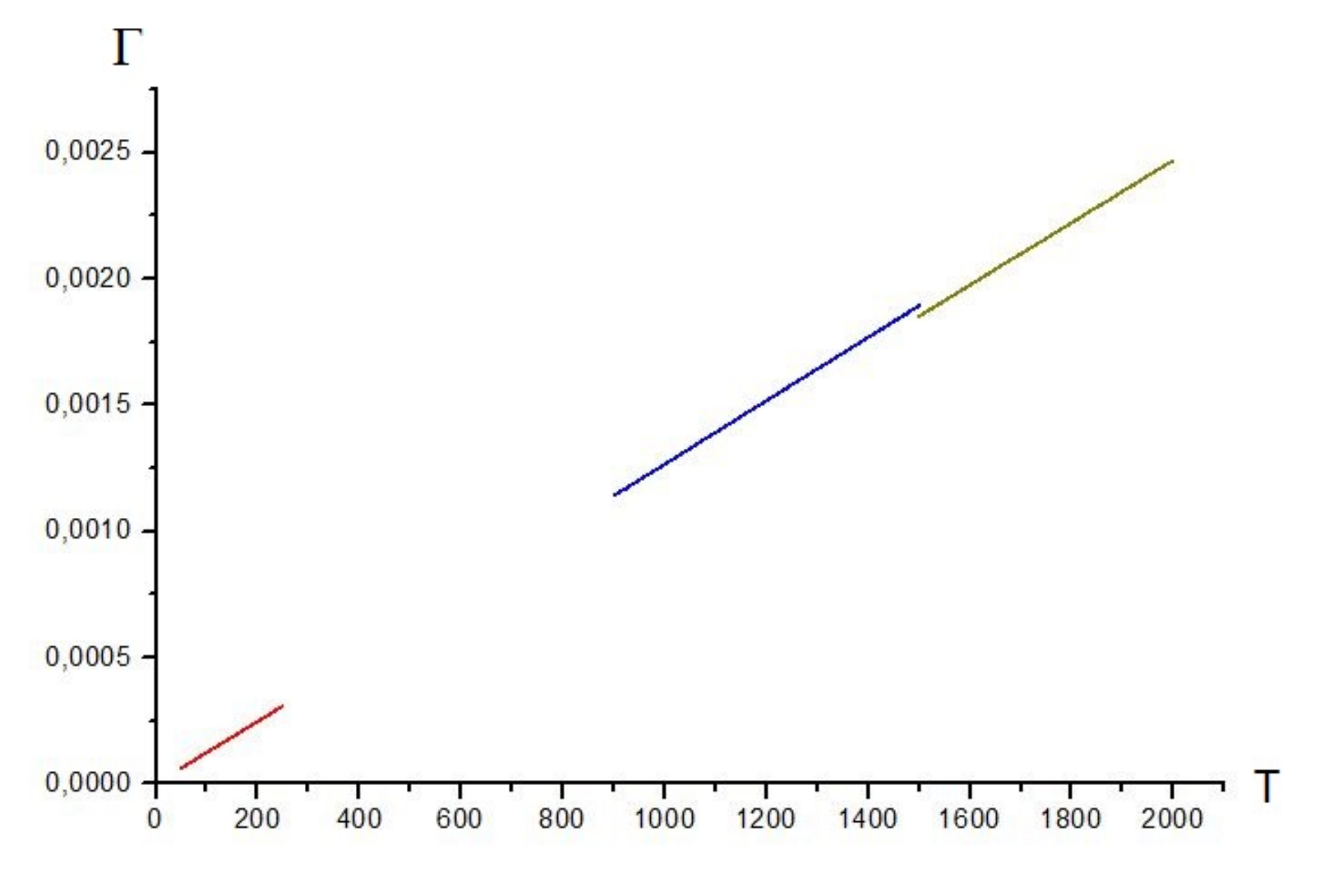} 
\end{center}
\caption{(color online only) The Sphaleron rate $\Gamma$ as a function of the temperature $T$ for the three steps of the phase transition.
}
\label{fig8}
\end{figure}
\item {Thin wall approximation}: Following ref. \cite{Sphaleron331}, we assume that,   
\begin{equation}
\frac{\partial V_{eff}(\upsilon _{\chi, \eta, \rho })}{\partial \upsilon _{\chi \eta, \rho  }}=C_{\chi, \eta, \rho}= const.
\end{equation}
where $\upsilon _{\chi, \eta, \rho }$ are the second minimum of the effective potential $V_{eff}$ in the bubble phase trantision
$SU(4)\rightarrow SU(3)$, $SU(3)\rightarrow SU(2)$ and $SU(2)\rightarrow U(1)$ respectively.
In this case the field equations of the VeVs $\upsilon _{\chi, \eta, \rho }$ read
\begin{equation}
\frac{d^{2} \upsilon_{\chi, \eta, \rho }} {dr^{2}}+\frac{2}{r}\frac{d\upsilon _{\chi, \eta, \rho }}{dr
} = C_{\chi, \eta, \rho} \label{eqm1} 
\end{equation}
with the boundary conditions
\begin{equation}
\underset{r\rightarrow \infty }{\lim }\upsilon _{\chi,  \eta, \rho }(r)=0, \,\,\,\,\,\,\,\,\,\,\,\,\,\,\,\,
\left. \frac{d\upsilon _{\chi,  \eta, \rho }(r)}{dr}\right\vert _{r=0}=0 
\label{ggg} 
\end{equation}
The solutions of eqs. (\ref{eqm1}) and (\ref{ggg}) are given by \cite{Sphaleron331}
\begin{equation}
\upsilon _{\chi, \eta, \rho } =\frac{C_{\chi, \eta, \rho }}{6}r^{2}-\frac{A_{\chi, \eta, \rho }}{r}+B_{\chi, \eta, \rho }  
\label{mab}
\end{equation}
where $A_{\chi, \eta, \rho}$, $B_{\chi, \eta, \rho }$ are integration constants. To be more specific if the sphaleron has a radius 
$R_{\chi, \eta, \rho }$ and a thickness  $\Delta l_{\chi, \eta, \rho }$ the solution $\upsilon _{\chi,  \eta, \rho }$ can be expressed 
as 
\begin{equation}
\upsilon_{\chi,  \eta, \rho }(r)=
\left\{ 
\begin{array}{lcr}
\upsilon _{\chi, \eta, \rho _{c}} \,\,\,\,\,\,\,\,\,\,\,\,\,\,\,\,\,\,\,\,\,\,\,\,\,\,\,\,\,\,\,\,\,\,\,\,\,\,\,\,\,\,\,\,\,\,\text{          when            
} r\leq R_{\chi, \eta, \rho} \\ 
\frac{C_{\chi,  \eta, \rho }}{6}r^2-\frac{A _{\chi,  \eta, \rho }}{r}+B _{\chi,  \eta, \rho }\text{   when       }R_{\chi,  \eta, \rho }
<r\leq R _{\chi,  \eta, \rho }+\Delta l _{\chi,  \eta, \rho } \\ 
0 \,\,\,\,\,\,\,\,\,\,\,\,\,\,\,\,\,\,\,\,\,\,\,\,\,\,\,\,\,\,\,\,\,\,\,\,\,\,\,\,\,\,\,\,\,\,\,\,\,\,\,\,\,\,\,\,\,\,\,\text{ when }R _{\chi,  \eta, 
\rho }+\Delta l _{\chi,  \eta, \rho }<r
\end{array}
\right.  \label{sol1}
\end{equation}
Here $\upsilon _{\chi, \eta, \rho _{c}}$ stands for the second minimum for the $3$ steps of the phase transition. In 
order to proceed further the constant $C_{\chi, \eta, \rho}$ can be approximated as 
\begin{equation}
C_{\chi, \eta, \rho}\sim \frac{\Delta V_{eff}(\upsilon _{\chi, \eta, \rho})}{\Delta \upsilon _{\chi, \eta, \rho}}
\end{equation}
where $\Delta V_{eff}= V_{eff_c}(\upsilon _{\chi, \eta, \rho _{c}})$ and $\Delta \upsilon _{\chi, \eta, \rho}=\upsilon _{\chi, 
\eta, \rho _{c}}$. Now, for the numerical results and in order to avoid the washout of the baryonic asymmetry after the phase 
transition 
one has to assume that the sphaleron rate $\Gamma$ has to be equal to the Hubble parameter $H$ at the critical 
temperature $T_c$. Of course $\Gamma$ has to be lager than $H$ at $T>T_c$ and smaller at $T<T_c$  see ref. 
\cite{Sphaleron331}, \cite{hubble 1}, \cite{hubble 2}.

Analyzing figures \ref{fig9} to \ref{fig11}  a general behavior was observed for all the EWPT $ SU(4)\rightarrow SU(3)$, 
$ SU(3)
\rightarrow SU(2)$ and $ SU(2)\rightarrow U(1)$: as $T$ decreases from the bubble nucleation temperature $T_1$ where 
the strong first-order phase transition starts the radius $R$ and the energy $\varepsilon$ of the bubble increase while the 
sphaleron rate $\Gamma$ decreases such that the ratio$\frac{\Gamma}{H}$ is bigger (resp. smaller) than $1$ for  
$T_c<T<T_1$ (resp. $T_0<T \leq T_c$).
To be more precise, we notice that the gauge symmetries respectively $ SU(4)
\rightarrow SU(3)$, $ SU(3)\rightarrow SU(2)$ and $ SU(2)\rightarrow U(1)$ start to  be broken spontaneously at the bubble 
nucleation temperature $T_1\sim 1823$, $1309.35$, $194.94$ GeV ($\hbar=c=k=1$). Then, a small bubble with radius 
$R\sim 13.53 \times 10^{-4}$, $4.5\times 10^{-5}$, $11.97\times 10^{-4}$ GeV$^{-1}$ and thickness $\Delta l\sim 10^{-4}$, 
$10^{-5}$, $10^{-4}$ (GeV$^{-1}$) appears and stores the nonvanishing VeVs $\upsilon _{\chi, \eta, \rho}$ inside. It is very 
important to mention that as it was pointed out in section \ref{EPT}, if the two minima are 
separated by a potential barrier the phase transition will occur with bubbles 
nucleations governed by thermal channeling from a local minimum at $\phi=0$ (false vacuum), to a deeper minimum at $
\phi\neq 0$ (true vacuum). The non-thermal equilibrium is induced by the rapidly expanding bubble walls through the 
cosmological 
plasma, and $B$-violation arises from the rapid sphaleron transition in the symmetrical phase. At this temperature, the 
sphaleron rate $\Gamma$ gets the 
values $4.8456\times10^{11}$, $2.8493\times10^{11}$, $3.11674\times10^{10}$ GeV, which are larger than the values of 
$H\sim 4.6769\times 10^{-12}$, $2.4126\times 10^{-12}$, $5.34795\times 10^{-12}$ GeV.
When the temperature decreases from $T_1$ $\sim1823$, $1309.35$, $194.94$ GeV to $T_c$ $\sim1820$, $1306$, $167.5$ 
GeV, the bubble volume and energy 
increases and decreases respectively. Moreover, the rate $\Gamma$ decreases but the ratio $\frac{\Gamma}{H}$ remain 
greater 
than $1$ allowing the bubbles to collide and fill all the space. This phenomenon is very violent leading to a huge deviation from 
thermal equilibrium. 
The baryon production takes place in the neighborhood of the expanding bubbles walls generating CP and C violation (which
 is not the scoop of our paper).
 In fact, for an illustration if $T=1822$, $1308$, $180$ GeV, $R=13.91\times 10^{-4}$, 
$9.5\times 10^{-5}$, $13.55\times 10^{-4}$, $\varepsilon=17398.496$, $24853.266$,  $3711.643$ GeV, $\Gamma= 3.7084\times 10^7$, 
$2091.05642$, $56.98872$ GeV and $\frac{\Gamma}{H}=7.9379\times 10^{18}$, 
$8.6848\times 10^{14}$, $1.2498\times 10^{15}$. Of course, as it was assumed before at $T=T_c$ the sphaleron rate $
\Gamma= H=4.66178 \times 10^{-12}$, $2.3913 \times10^{-12}$, $3.94764 \times10^{-14}$GeV. When $T$ becomes 
smaller than $T_c$ 
($T=1811.5$, $1300$, $158$ GeV) $\Gamma$ decreases rapidly and the ratio $\frac{\Gamma}{H}$ becomes less than $1$ 
($\frac{\Gamma}{H}
=3.19962\times 10^{-210}$, $3.2444\times 10^{-19}$, $1.8254\times 10^{-109}$). As the temperature reaches the 
transition ending temperature 
$T_0= 1796.625, 1280.02, 151.89$ GeV, only the broben phase remains and the sphaleron transitions $ SU(4)\rightarrow 
SU(3)$, $ SU(3)
\rightarrow SU(2)$ and $ SU(2)\rightarrow U(1)$ are totally shut off. 

\end{enumerate}

\begin{figure}[t]
\begin{center}
\begin{tabular}{cccccccccccccccc}
\includegraphics[scale=0.272]{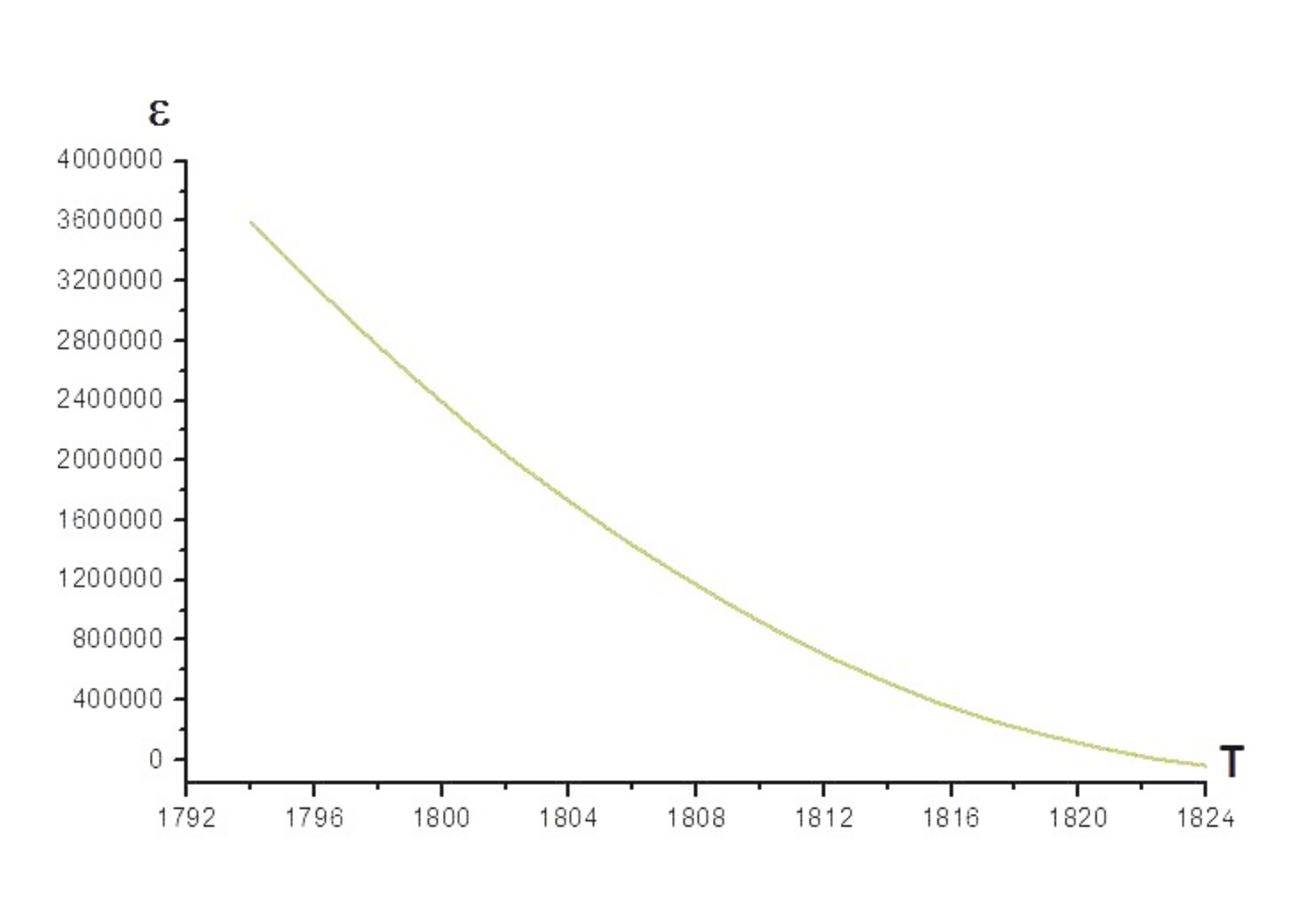} 
&
\includegraphics[scale=0.272]{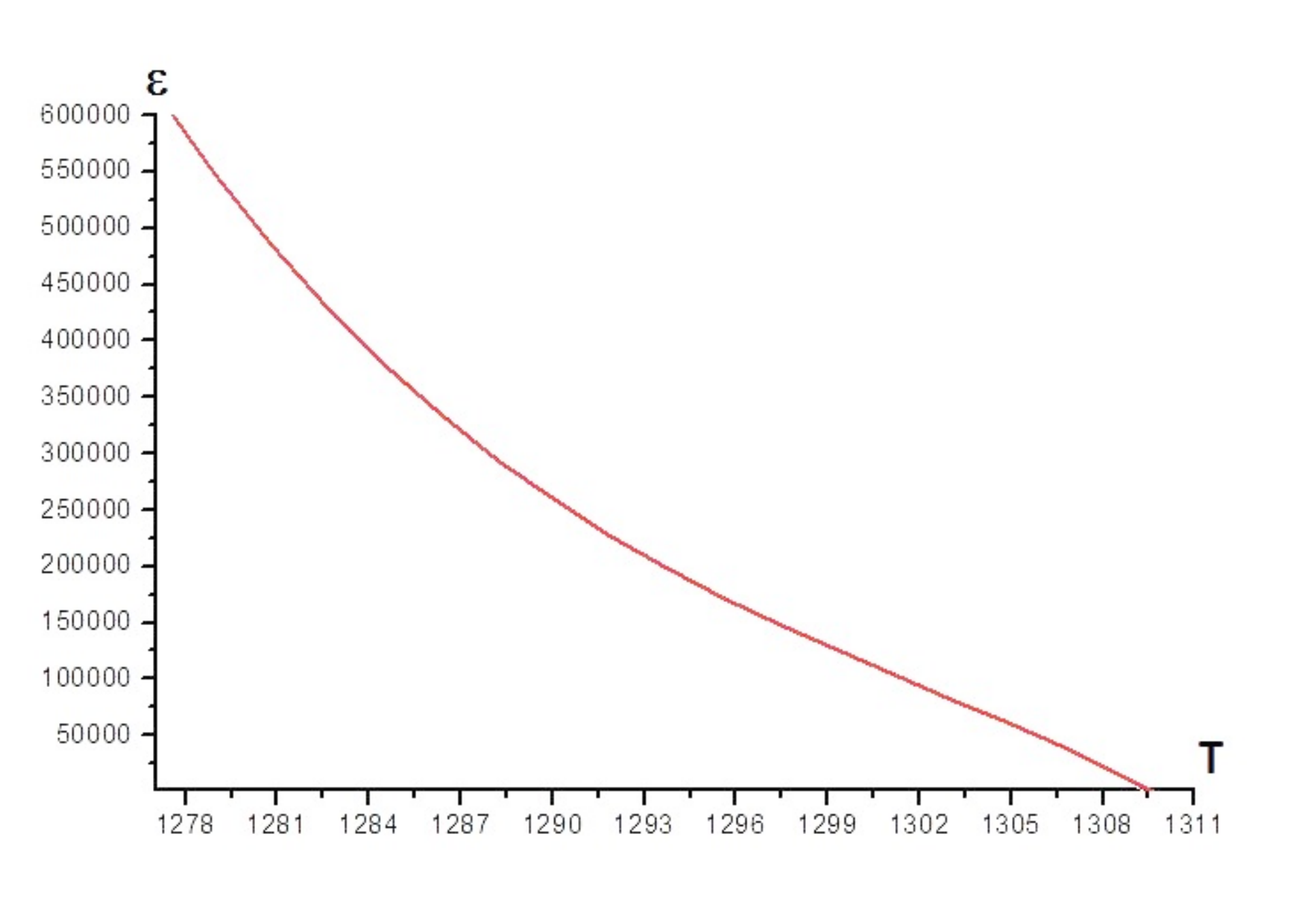} 
&
\includegraphics[scale=0.272]{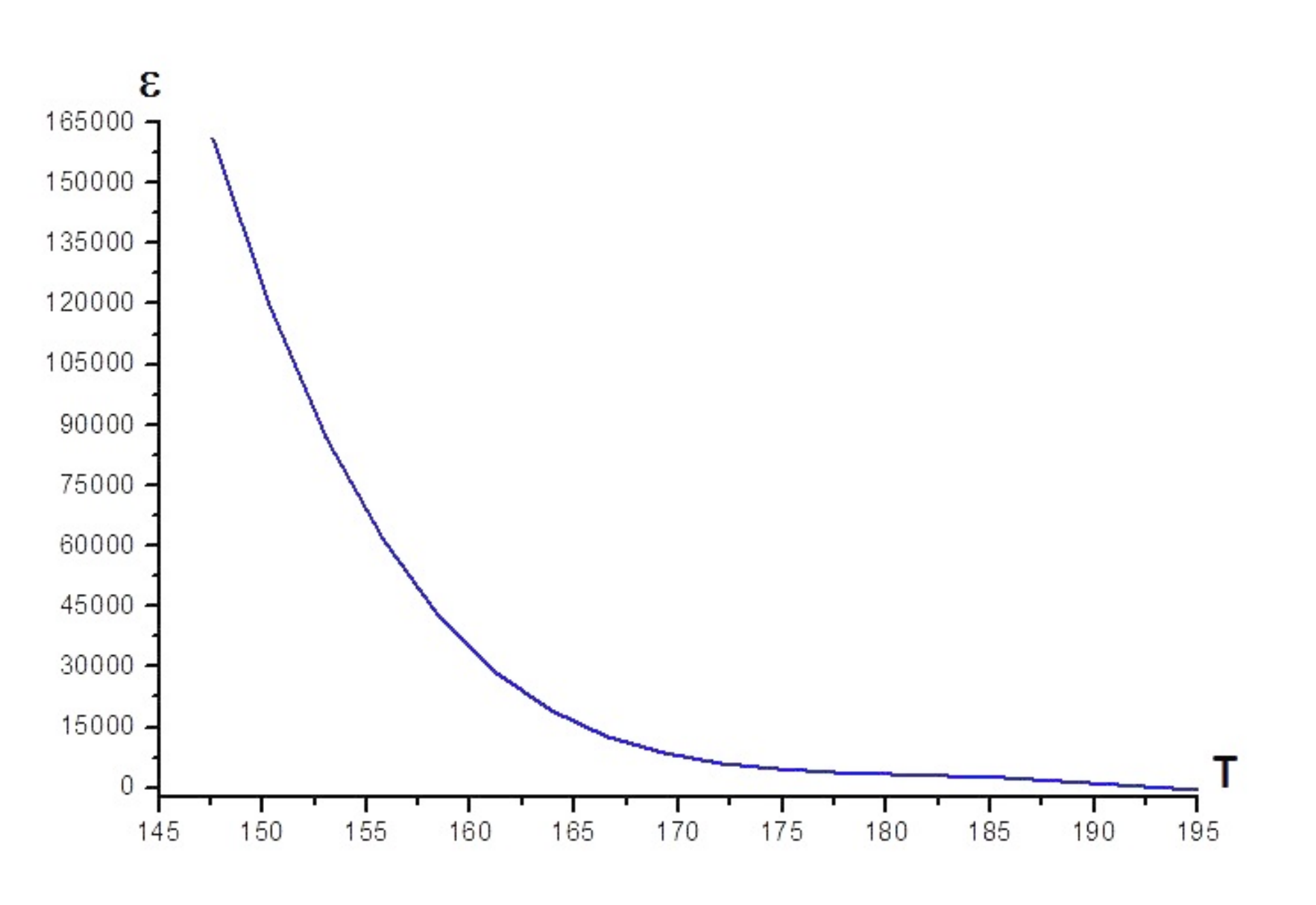} 
\end{tabular}
\end{center}
\caption{(color online only) The sphaleron energy $\varepsilon$ as a function of the temperature $T$ for the three steps of the phase transition.}
\label{fig9}
\end{figure}
\begin{figure}[t]
\begin{center}
\begin{tabular}{cccccccccccccccc}
\includegraphics[scale=0.273]{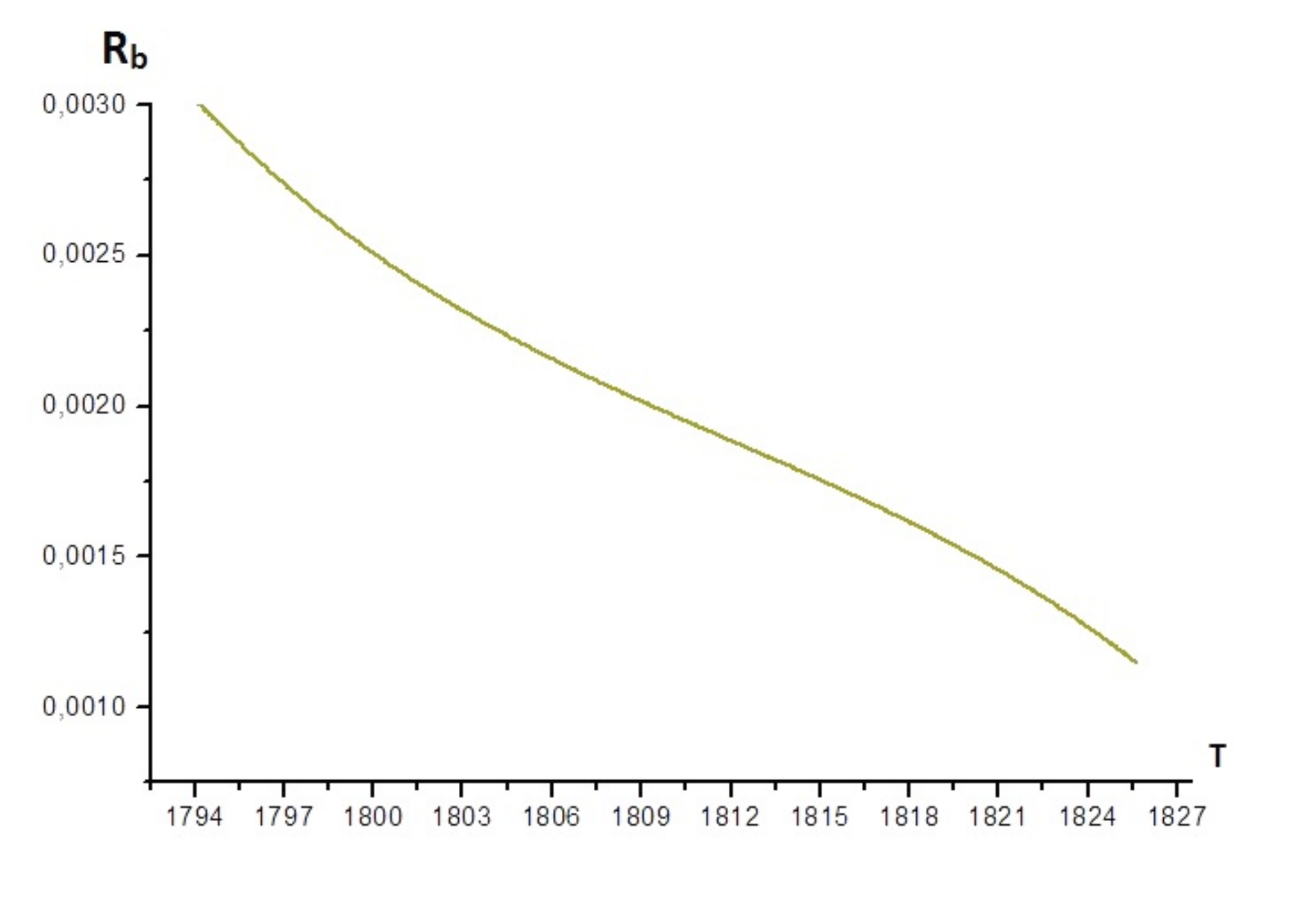} 
&    
\includegraphics[scale=0.273]{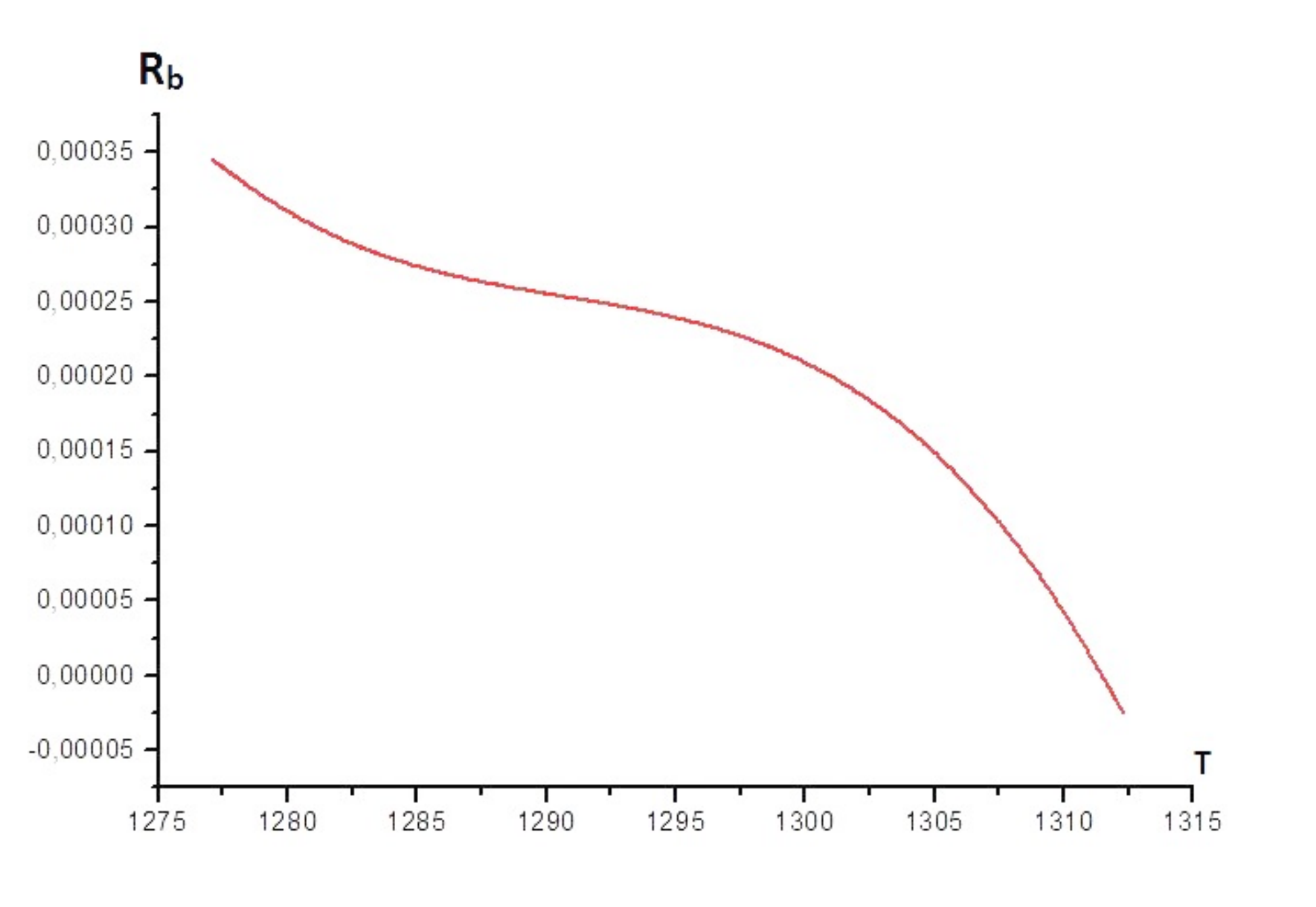} 
& 
\includegraphics[scale=0.273]{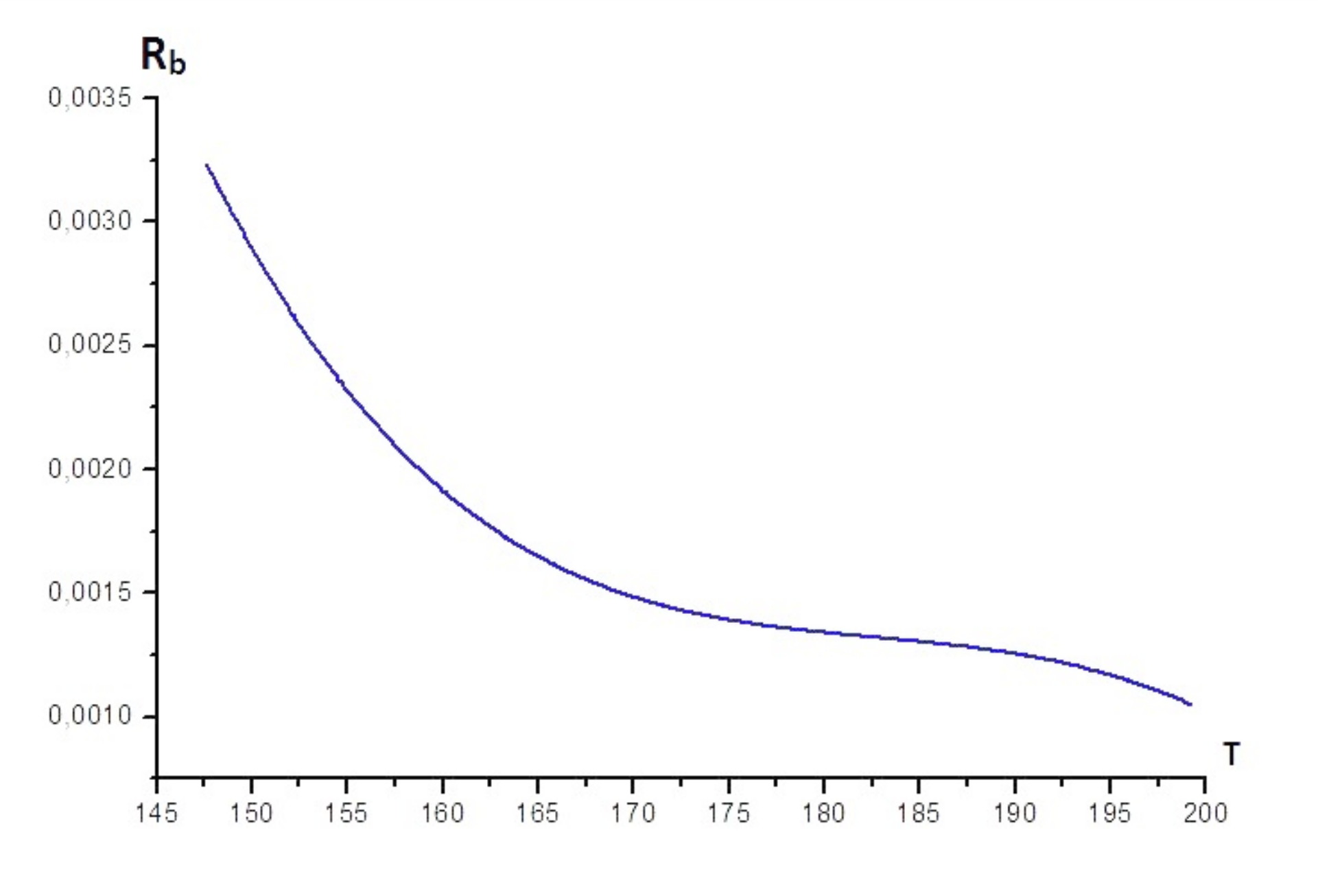} 
\end{tabular}
\end{center}
\caption{ (color online only) The radii of the bubbles $R$ as a function of the temperature $T$ for the three steps of the phase transition.}
\label{fig10}
\end{figure}
\begin{figure}[t]
\begin{center}
\begin{tabular}{cccccccccccccccc}
\includegraphics[scale=0.27]{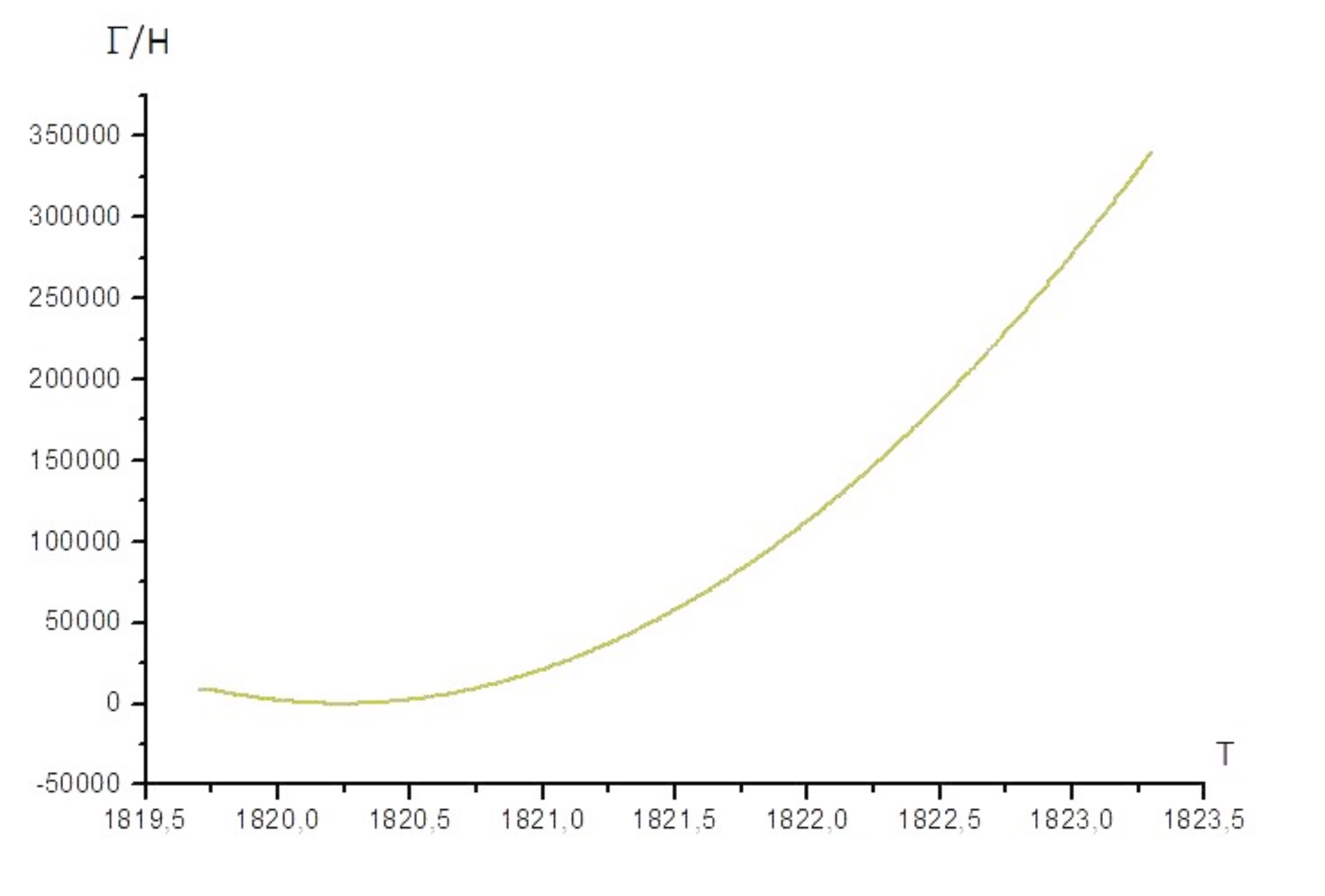} 
&
\includegraphics[scale=0.27]{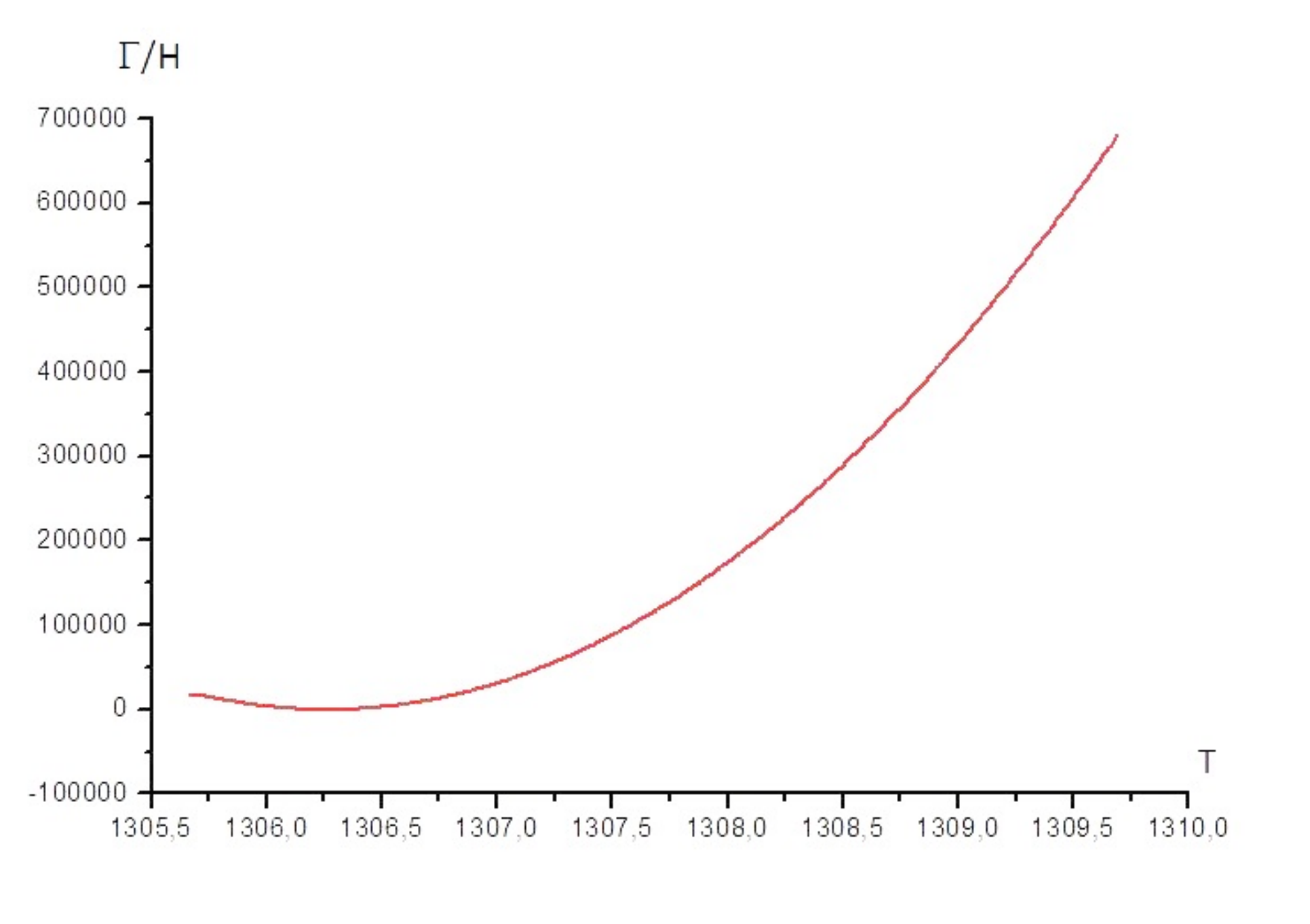} 
& 
\includegraphics[scale=0.27]{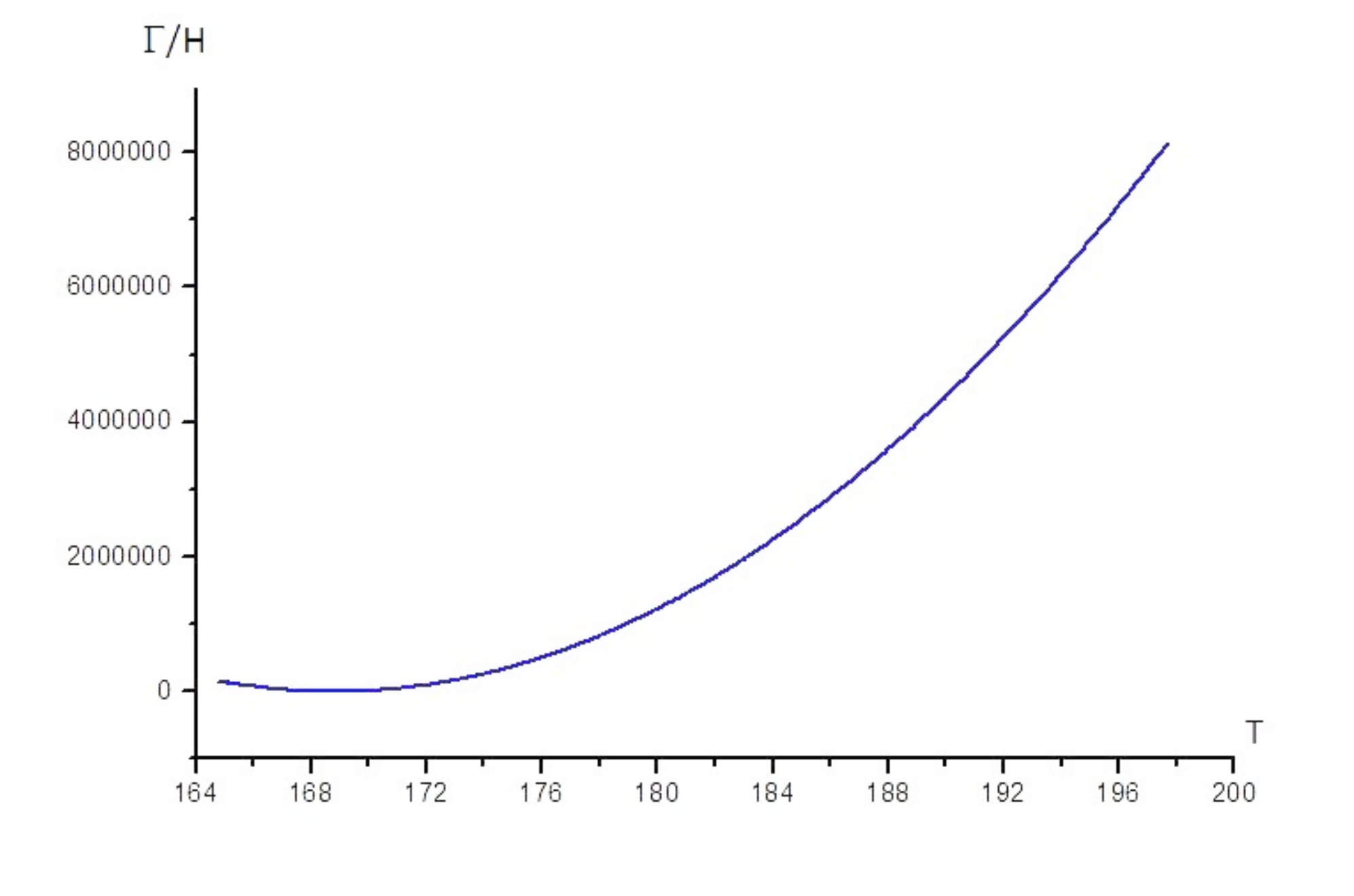} 
\end{tabular}
\end{center}
\caption{(color online only) The ration $\frac{\Gamma}{H}$ as a function of the temperature $T$ for the three steps of the phase transition.}
\label{fig11}
\end{figure}

\section{Conclusion}
\label{sect6}
Throughout this paper, and in order to be self-consistent, theoretical constraints on the potential parameters 
such as the unitarity, stability, and boundness have been imposed. Moreover, using a Monte-Carlo simulation, 
we have bound the various allowed regions of the parameter space verifying the first-order phase transition criteria 
at  $\upsilon _{\chi }\sim \upsilon_{\eta }\sim 2\ $ TeV and $\upsilon _{\rho }=246$ GeV for the three steps 
$ SU(4)\rightarrow SU(3)$, $ SU(3) \rightarrow SU(2)$, and $ SU(2)\rightarrow U(1)$  respectively, leading to 
an effective potential and confidence bands where masses of the heavy Higgs 
bosons are in the range of $700-1300$ GeV. Moreover, we have derived the expressions of the effective potential,
nucleation, critical, and ending temperatures in terms of particle masses and temperature for each step of the EWPT. 
Furthermore, the baryogenesis study using the sphaleron approach was also investigated, where it is shown that the static 
approximation cannot give consistent results for the ratio $\frac{\Gamma}{H}$ for $T<T_c$ why the thin wall approximation 
does. Finally, we can conclude that we have obtained the same conclusions as those of ref.\cite{Sphaleron331} but within the framework
 of the $331$  model. It is very important to stress out that, 
the authors of ref.\cite{Sphaleron331} did not impose the theoretical constraints on the potential parameters as we did. 
Further investigations on CP violation in this model are under consideration.


\section*{Acknowledgments}
The authors would like to thank the Algerian Ministry of Higher Education and Scientific Research as well as the DGRSDT for financial supports.

\end{document}